\documentclass{article}
\usepackage[margin=1in]{geometry}
\usepackage{graphicx} 
\usepackage{amssymb,amsmath,color}
\usepackage{hyperref}
\hypersetup{colorlinks=true, citecolor=red, urlcolor=blue, linkcolor=blue}
\usepackage{caption}
\usepackage{physics}
\usepackage{authblk}

\newcommand{\name}{Spin SYK}
\renewcommand{\t}{\tau}
\newcommand{\q}{q_{\textrm{EA}}}
\newcommand{\av}[1]{\mathbb{E}\left[ #1 \right]}

\newcommand{\gsyk}{$q$-local Gauged Clusters}

\newcommand{\CD}{\mathcal{D}}

\title{A Bosonic Model of Quantum Holography}
\author[1]{Brian Swingle} 
\author[2]{Mike Winer}
\affil[1]{Department of Physics, Brandeis University, Waltham, Massachusetts 02453, USA}
\affil[2]{Joint Quantum Institute, Department of Physics, University of Maryland, College Park, Maryland 20742, USA}
\date{\today}

\begin{document}

\maketitle
\begin{abstract}
    
    We analyze a model of qubits which we argue has an emergent quantum gravitational description similar to the fermionic Sachdev-Ye-Kitaev (SYK) model. The model we consider is known as the quantum $q$-spin model because it features $q$-local interactions between qubits. It was previously studied as a model of a quantum spin glass, and while we find that the model is glassy for $q=2$, $q=3$, and likely $q=4$, we also find evidence for previously unexpected SYK-like behavior for the quenched free energy down to the lowest temperatures for $q \geq 5$. This SYK-like physics includes power-law correlation functions and an extensive low temperature entropy, so we refer to the model as \name{}. The model is generic in that it includes all possible $q$-body couplings, lacks most symmetries, and has no spatial structure, so our results can be construed as establishing a certain ubiquity of quantum holography in systems dominated by many-body interactions. Furthermore, we discuss a generalized family of models which includes \name{} and which provably exhibit SYK-like physics in the solvable limit of large local Hilbert space dimension. We also comment on implications of a bosonic system with SYK-like properties for the study of holography, Hamiltonian complexity, and related topics. 

\end{abstract}

\section{Introduction}

In this paper we study the low-energy physics of generic $q$-local quantum systems. In their simplest form, these are Hamiltonians that include all possible $q$-body interactions between $N$ sites with minimal additional structure. The models we consider are bosonic in that their Hilbert space has a tensor product structure built from the sites such that operators from different sites commute. Variants of such ``mean-field'' models have a long history, including as models of quantum spin glasses \cite{Bray_1980,SachRead,Sachdev_1994,christos_2022,georges_2000,Biroli_2001,anous_2021,erdos_pspin}, as models for investigating Hamiltonian complexity~\cite{sevag_ham_complexity,harrow_product,keating_spectra,chifang_sparse_easy}, and, when generalizing to fermions, as simple models of holography \cite{SY,kitaev2015simple,Maldacena_2016,Rosenhaus_2019,Davison_2017,Gaikwad_2020,kobrin2020manybody,Sorokhaibam_2020,Fu_2017,Sun_2020,xu2020sparse}. Most of this paper is concerned with what we call the \name{} model, for which the sites are qubits. We argue that \name{} is glassy at low energies when $q=2,3,4$ but at larger $q$ exhibits holographic dynamics similar to the fermionic Sachdev-Ye-Kitaev model, hence the moniker. Even for smaller $q$ for which the physics appears to be glassy at the lowest energies, \name{} can still exhibit SYK-like physics over a window of energies. Observations of SYK-like physics at intermediate energies has also been reported in the $q=2$ random Heisenberg model~\cite{shackleton_jt,haehl_crossover}.

We also discuss a generalization of \name{}, the \gsyk{} model, in which the qubits are replaced by sites with local Hilbert space dimension equal to a power of $2$ and with interactions drawn from a large but proper subset of all local operators. The model is designed to be similar to random SU($M$) quantum magnets, which are solvable at large $M$~\cite{SY}. In the limit of large local Hilbert space dimension, this \gsyk{} also becomes solvable and is SYK-like for any $q\geq 2$, although it suffers from a known low-temperature instability~\cite{Bray_1980,SY,haehl_crossover} towards glassiness when $q=2$ and the local Hilbert space is finite.  

Our results indicate that the holographic low-energy physics of the Sachdev-Ye-Kitaev (SYK) model is actually more generic then previously known. Compared to \name{}, the presence of low temperature holographic dynamics appears to be only slightly more robust in SYK, which is not glassy at $q=4$~\cite{gurari_syk_glass} and for which only $q=2$ is non-holographic. The SYK model exemplifies many features of strong coupling dynamics and provides a simple realization of holographic duality~\cite{maldacena_adscft}, and these physical properties, combined with its analytical tractability, have led to its intense study. Compared to SYK, the bosonic models we consider are interesting for several reasons: (1) they might be easier to implement in quantum simulations, especially in the near term, (2) one can potentially get SYK-like physics at intermediate energies even when $q=2$, and (3) they provide models of holography in which all the microscopic degrees of freedom are bosonic. \name{} is also interesting for its potential applications in Hamiltonian complexity, e.g.~\cite{chifang_sparse_easy}, and as a launching point for further model development. Finally, \name{} has the interesting property that SYK dynamics appear at a quenched saddle point, the one which dominates the dynamics and thermodynamics for a random couplings. The annealed saddle point, which dominates the disorder-averaged partition function, is still glassy. This is in contrast to the SYK model, where the quenched and annealed saddle points are the same.

As we have said, the bulk of the paper studies the \name{} model, which was introduced in \cite{erdos_pspin} (where it was called a quantum $p$-spin glass model) and further studied in \cite{berkooz_chord_sg,Baldwin_2020,Hanada:2023rkf,anschuetz2023product}, with Hamiltonian given by
\begin{equation}
    H = \sum_{r_1\mu_1 \cdots r_q \mu_q} J_{r_1 \mu_1 \cdots r_q \mu_q} \sigma_{r_1 \mu_1} \cdots \sigma_{r_q \mu_q}
    \label{eq:SpinSYKHam}
\end{equation}
where $r=1,\cdots, N$ refers to the site and $\mu=1,\cdots,3$ refers to the spin component (Pauli matrix). The $J$ couplings are Gaussian random variables with zero mean and variance
\begin{equation}
    \av{J_{r\mu \cdots}^2}=\frac{(q-1)! J^2}{N^{q-1}}.
\end{equation}
We highlight in particular the recent independent study~\cite{Hanada:2023rkf} which considered a variant of \eqref{eq:SpinSYKHam} with $q=4$ and only two Pauli components $\sigma_x,\sigma_y$ instead of all three and gave evidence from finite size numerics that the physics was SYK-like over a wide range of energies. As we discuss in Section~\ref{sec:discussion}, many of our conclusions will apply to the two component case as well.

The main assertion of this paper is that model \eqref{eq:SpinSYKHam} displays power-law decay in its low temperature correlation functions for sufficiently large $q$. In the rest of this section, we will define the model and the primary observables of interest. We will also discuss what we mean when we calculate the disorder-averaged properties of this model and overview our analysis strategy.

In Section \ref{sec:Action} we will write down the action for a path integral formulation of the \name{} thermal partition function in terms of a correlation function $G$ and a self-energy $\Sigma$. We work in the large $N$ limit in which the path integral can be evaluated using saddle-point methods. We will differentiate the \name{} action to get saddle-point equations for $G$ and $\Sigma$.

In Section \ref{sec:Solutions} we will discuss possible solutions of these equations. We will prove the existence of two saddle-points of the single-replica action: an annealed solution with $G\to 1$ at low temperatures which dominates the naive average of the partition function, and a possible quenched solution which could control the free energy one would actually measure. This solution is replica-diagonal and gapless, and obeys a power-law decay. 

Section \ref{sec:Liouville} examines the large-$q$ limit of the \name{} model, and shows that it obeys a Liouville equation just like the traditional SYK model. This means that the large $q$ properties are essentially identical to that of the SYK model.

Section \ref{sec:Numerics} studies numerical solutions to the single-replica saddle-point equations for general $q$. For $q=2$, we find only the annealed solution. For $q>2$, we find the annealed solution and another solution which we refer to as SYK-like. Moreover, even as low as $q=3$, the SYK-like solution to the Liouville equations is a good qualitative fit for (an) exact solution. We also find that the SYK model's conformal solution is a good fit for the SYK-like solution at low temperature.

Section \ref{sec:ed} goes over exact-diagonalization results, which suggest that the system is glassy at $q=2,3,4$ and SYK-like for $q\geq 5$. We study a sparse version of \name{} which allows us to address larger $q$ efficiently. Our main strategy is to measure the Edwards-Anderson order parameter. When it is zero, then the system should be replica diagonal and the power-law solution dominates. When it is non-zero, the system is presumably glassy. We also look at low-energy level spacings, finding results consistent with SYK-like physics and inconsistent with a spin glass for $q\geq 5$.

Section \ref{sec:q_col} deals with a generalization of \name{} in which the qubits are replaced by sites of Hilbert space dimension $ 2^{M/2-1}$ for an even integer $M\geq 4$. These sites have an interpretation as $M$ Majorana fermions, with the local fermion parity gauged to produce a bosonic local Hilbert space, hence \gsyk{}. In this picture, we make a set of all possible fermion bilinears (which are bosonic operators) and consider generic $q$-local interactions built from these operators. \name{} is the special case $M=4$. At large $M$, the model is solvable and exhibits SYK-like physics over a wide temperature range for any $q \geq 2$. We conjecture that this behavior persists all the way down to $M=4$ for sufficiently large $q$.

In Section \ref{sec:syk-like} we elaborate on just how similar the low energy physics is between SYK and \name{}. We discuss the ground state entropy, the emergence of spontaneously and explicitly broken time-reparameterizations, the Schwarzian action, and the possible effects of weakly irrelevant operators.

Finally, in Section \ref{sec:discussion} we conclude with a brief outlook that highlights many questions remaining to be addressed. A few appendices contain the derivation of the \name{} action and details of the numerical calculations. 

This paper represents the first part of an ongoing investigation. We plan to report on a detailed study of multi-replica saddles in a future work.

\subsection{Model and Observables}

We consider a model defined on $N$ spin-1/2 degrees of freedom. We also fix an integer $q>1$, the ``locality'', which determines the degree of the interaction. The terms in the Hamiltonian are obtained by considering all possible $q$-body subsets of the spins and an assignment of $\sigma_\mu$ ($\mu=1,\cdots,3$) for each spin. This yields an ensemble of Hamiltonians where each instance takes the form  
\begin{equation}
    H = \sum_{r_1\mu_1 \cdots r_q \mu_q} J_{r_1 \mu_1 \cdots r_q \mu_q} \sigma_{r_1 \mu_1} \cdots \sigma_{r_q \mu_q} \label{eq:HamAgain},
\end{equation}
with $r=1,\cdots, N$ referring to the spin and $\mu=1,2,3$ referring to the component. The model does not have spin-rotation symmetry (except in a statistical sense), and we typically use spin and qubit terminology interchangeably.

Our goal in this paper is to analyze the Hamiltonian ensemble defined in \eqref{eq:HamAgain}. Our primary interest is to obtain the energy and entropy in the Gibbs state $e^{-\beta H}$ as a function of $\beta$, averaged over realizations of the $J$ couplings. We focus on the low temperature regime of $N\to \infty$ and $\beta J$ large.

Crucial to our considerations will be the Green's function or correlation function,
\begin{equation}
    G(\tau_1,\tau_2) = \frac{1}{N} \sum_{r\mu} \av{\langle \sigma_{r\mu}(\tau_1) \sigma_{r\mu}(\tau_2) \rangle_{\beta}},
\end{equation}
where we take both a thermal and an ensemble average. The arguments of the Pauli matrices indicate imaginary time values, with $O(\tau) = e^{\tau H} O e^{-\tau H}$ for $\tau \in [0,\beta)$. As we will see below, the low-temperature fate of the model is closely tied to the properties of this correlation function. In our work, $G(\t_1,\t_2)$ will be translation invariant, so that $G(\t_1,\t_2)=G(\t_1-\t_2,0) \equiv G(\t_1-\t_2)$.

At very large $\beta$, an important role is played by the asymptotic value $G(\infty)$. From the definition of $G$, this is equal to the ensemble average of the ``Edwards-Anderson'' order parameter in the ground state,
\begin{equation}
    G(\infty) = \frac{1}{N} \sum_{r\mu} \langle \text{gs} | \sigma_{r \mu} | \text{gs} \rangle^2 \equiv q_{\text{EA}}.
\end{equation}
This notation is conventional (although our normalization of $\q$ is slightly different); $q_{\text{EA}}$ has nothing directly to do with $q$, the locality. A non-zero value of $q_{\text{EA}}$ indicates a frozen pattern of local expectation values in the ground state, which is a hallmark of a glass. In contrast, when $q_{\text{EA}}$ is zero at $\beta=\infty$, at least the equilibrium thermodynamics of the model is not glassy. An important implication of $q_{\text{EA}} >0$ is that if we have multiple ``replicas'' of the system with identical $J$ couplings, then the frozen pattern of expectation values will have some similarity or ``overlap'' between different replicas. This overlap is also controlled by $q_{\text{EA}}$ at large $\beta$.

\subsection{Quenched vs Annealed}

We are interested in the disorder-averaged free energy, $F_q$ \cite{foini_2022}. This is a ``quenched'' average, meaning we compute the free energy for a given sample (a realization of the $J$ couplings), and then average over samples,
\begin{equation}
    F_q \equiv \int dJ P(J) F(J) \equiv \av{F}.
\end{equation}
Since $F$ is related to the partition function $Z = \tr(e^{-\beta H})$ via
\begin{equation}
    F = - \frac{1}{\beta} \ln Z,
\end{equation}
this quenched average corresponds to averaging $\ln Z$,
\begin{equation}
    F_q = -\frac{1}{\beta} \av{\ln Z}.
\end{equation}

As is well known, it can be challenging to compute the average of $\ln Z$. The standard method is to use a replica trick in which we introduce $n$ copies or replicas of the system, compute the average of $Z^n$, and then take $n 
\to 0$ using the identity
\begin{equation}
    \av{\ln Z} = \lim_{n\to 0} \frac{\av{Z^n}-1}{n}.
\end{equation}
Although the replica trick is known or conjectured to have issues in some cases~\cite{Tanaka2007Moment,Mourrat2021Nonconvex}, we will assume it is valid and that we can exchange the order of limits of $N \to \infty$ and $n \to 0$.

Even with this assumption, evaluating the ``replica limit'' $n \to 0$ is still typically challenging. Our approach is to express the partition function as a path integral. Then the disorder average of $Z^n$ is straightforward to carry out and we are left with a ``replicated'' path integral to evaluate. In the large $N$ limit, we write this replicated path integral schematically as
\begin{equation}
    \av{Z^n} = \int \CD \phi e^{- N I_n[\phi]}
\end{equation}
for some general integration variables $\phi$ and ``replicated action'' $I_n$. Taking the large $N$ limit suggests looking among the saddle points of $I_n$ for the physical saddle. However, whereas one might naively look for saddles of smallest action, there is a well known prescription due to Parisi that states the correct answer in the replica limit is actually to look for saddles that are maximal~\cite{parisi_max}, at least with respect to ``replica order parameters''. See Eq. \eqref{eq:ssyk_rep_action} for details of the replicated action for \name, written in terms of a variable $G$ which will reduce to the physical correlation function at the physical saddle point and a corresponding self-energy $\Sigma$, which jointly play the role of $\phi$. 

A far simpler quantity to compute is the logarithm of the average of $Z$. This order of operations defines the ``annealed'' free energy, 
\begin{equation}
    F_a = - \frac{1}{\beta}\ln \av{Z},
\end{equation}
and for SYK we have the remarkable fact that
\begin{equation}
   \lim_{N \to \infty} \frac{F_q}{N} = \lim_{N \to \infty} \frac{F_a}{N}.
\end{equation}
This is an enormous technical simplification in that we only ever have to consider a single replica, at least for thermodynamic quantities. The way this works for SYK is that the relevant saddle point of the replicated action, $\phi_*$, has no correlation amongst the replicas. In this case, $I_n(\phi_*) = n I_1(\phi_*)$ and the replica limit can be taken without trouble. Thus, the quenched free energy is determined from $I_1(\phi_*)$. In this situation, we say the relevant saddle is ``replica diagonal'' (no correlations between replicas) thus ``replica symmetric'' (the $S_n$ symmetry among the replicas is unbroken). 

Unfortunately, it was recently shown that the quenched and annealed free energies must differ for a large class of bosonic models below a model-dependent (but order one) temperature~\cite{Baldwin_2020}. The \name{} model satisfies the conditions of this theorem, so $F_q \neq F_a$ at sufficiently low temperature. In general, for systems like the \name{} model, at low temperatures the annealed free energy will linear in $\beta$, while the quenched free energy will be the ground state energy, which is independent of $\beta$. This implies that different saddle points compute the annealed and quenched free energies. The annealed saddle is given by a function $G_a(\t)$ which goes to some non-zero constant for large $\t$, a constant which is not directly related to $q_{\text{EA}}$. The quenched saddle $G_q(\t)$ must instead approach $\q$, the overlap between replicas. If this $\q$ is zero, then the quenched solution is replica diagonal and goes to zero at long times.  

This multitude of solutions does not imply that the model is necessarily glassy, although one often thinks of glassiness when quenched and annealed energies differ. But it does mean that we must deal with the replicated action and try to determine the correct saddle point, even if the correct solution is ultimately replica diagonal. We employ a recently proposed ``minimax'' procedure which extends Parisi's proposal~\cite{Baldwin_2023}. This procedure calls for first maximizing $I_n$ over replica order parameters (in this case the correlations between replicas) and then minimizing it over conventional order parameters (in this case the single-replica correlation). 
\subsection{Analysis Strategy}

In principle, the procedure is clear: write down the replicated action, find its saddle points, and carry out the minimax procedure to determine which saddle controls the quenched free energy. This calculation is possible but somewhat involved, so we are deferring it to future work. Here, we follow a different strategy in which we use finite size numerical results to tell us whether the quenched saddle is replica diagonal or not. When it is, we can obtain the quenched free energy from single-replica saddles, which we analyze in detail.

The key point is that, for disordered systems which are not in a glassy phase, for example, the Transverse Field Sherrington Kirkpatrick (TFSK)~\cite{USADEL1987975,PhysRevB.41.4858} model at large transverse field, it can be that the minimax saddle is replica diagonal but is distinct from the annealed saddle point (which is also replica diagonal). In this case, the minimax saddle can be obtained from the single replica equations of motion. It will always have higher action compared to the annealed saddle at sufficiently low temperature, nevertheless, the minimax procedure instructs us to take the minimax saddle over the annealed saddle when computing the quenched free energy.

Our strategy thus has two parts. First, we construct an inventory of possible single-replica saddles. This turns out to be a small set, containing just the annealed saddle and another SYK-like power-law saddle provided $q>2$. This is the business of Sections~\ref{sec:Action},\ref{sec:Solutions} \ref{sec:Liouville}, and \ref{sec:Numerics}. Second, we use exact diagonalization to directly study the overlap $\q$ in the ground state of sparse finite $N$ instances of the model. This provides an indirect way to determine whether the saddle which controls the quenched free energy is replica diagonal or not, although it relies on an extrapolation to large $N$. This is the business of Section~\ref{sec:ed}.

\begin{figure}
    \centering
    \includegraphics[width=.9\textwidth]{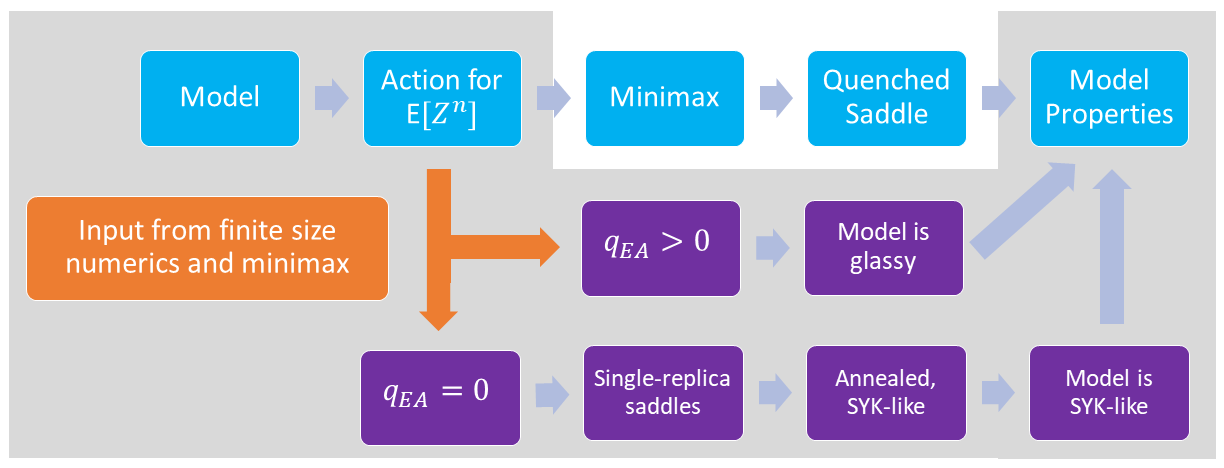}
    \caption{The strategy we follow in this paper is in grey. The top flowchart of blue boxes is the general procedure proposed in~\cite{Baldwin_2023} in which we construct the replicated action, carry out the minimax procedure, and then use the quenched saddle to determine the free energy. While this procedure should be possible to carry out, here we adopt an alternate procedure in which we try to determine $\q$ using an extrapolation from finite size numerics (orange box). When $\q=0$, we can restrict to single-replica saddles in which case our results indicate that the only possibility is the SYK-like saddle.}
    \label{fig:strategy}
\end{figure}

If $\q$ is zero in the thermodynamic limit, then whatever saddle is chosen by minimax at large $\beta$ must be replica diagonal. Since there is only one replica diagonal option (remembering that the annealed solution cannot compute the quenched free energy), we know that it must be the SYK-like saddle which controls the quenched free-energy. If $\q$ is non-zero in the thermodynamic limit, then the model is glassy. Our strategy is summarized in Figure \ref{fig:strategy}.

The main failure modes of this approach are two-fold. First, we might miss a relevant single-replica saddle. As we discuss below, we could find no other solutions and rule out large classes of solutions including exponentially decaying ones, but we cannot at this time rigorously prove that there are no additional solutions. Second, the sparse finite $N$ results might be misleading, either because of the finite sparsity or the finite system size. For $q\geq 5$, we observed an exponential decrease of $\q$ with system size and saw no evidence of significant deviations from this behavior, but cannot rule out a surprise reversal at large $N$.

\section{The Action of \name}
\label{sec:Action}
We begin our analysis by using spin coherent states \cite{annacoherent,SHIBATA_1999} to develop a path integral expression for disorder-average of the partition function $\av{Z}$. For the spin-1/2 Pauli matrices in the \name{} Hamiltonian, we have the map $\sigma_\mu \rightarrow 3 \Omega_\mu $ where $\Omega_\mu$ is a unit vector labeling a spin coherent state. We include all the overlaps of the spin coherent states into a measure $\CD^N \Omega$, c.f. \eqref{eq:measure_app}. The single replica average partition function is then
\begin{equation}
    \av{Z} = \int \CD^N \Omega \exp\left( \frac{\av{J_{r\mu \cdots}^2}}{2 q!}\sum_A \left[ \int_0^\beta d\tau s_{r_1 \mu_1}(\tau) \cdots s_{r_q \mu_q}(\tau) d\t\right]^2 \right),
\end{equation}
where $s_{rx}(\tau) = 3 \sin \theta_r(\t)\cos \phi_r(\tau)$, $s_{ry}(\t)=3 \sin \theta_r(\t)\sin \phi_r(\tau)$ and $s_{rz}(\t)=3 \cos \theta_r(\tau)$. This sort of path integral was written down for several closely related models in \cite{Baldwin_2020}.

We next introduce $G,\Sigma$ variables with the $\Sigma$ variable enforcing the constraint that
\begin{equation}
    G(\tau_1,\tau_2) = \frac{1}{N} \sum_{r\mu} s_{r\mu}(\tau_1) s_{r\mu}(\tau_2).
\end{equation}
This will correspond to the correlation function $\langle \sigma_x(\tau_1) \sigma_x(\tau_2) \rangle + \langle \sigma_y(\tau_1) \sigma_y(\tau_2)\rangle+ \langle \sigma_z(\tau_1) \sigma_z(\tau_2)\rangle$, averaged over sites. Up to $1/N$ corrections, the path integral is
\begin{equation}
    \av{Z} = \int \CD^N \Omega \CD \Sigma \CD G \exp\left(N\frac{J^2}{2 q} \int G^q d\t^2- \frac{N}{2} \int \Sigma \left[ G - \frac{1}{N} \sum_{r\mu} s_{r\mu}(\tau_1) s_{r\mu}(\tau_2) \right]d\t^2 \right).
    \label{eq:GSigma}
\end{equation}

The resulting integral is still hard to solve because of the spin bilinear. So we take one more step of introducing fluctuating magnetic fields $h_{x,y,z}(\tau)$, one for each spin, each with covariance $\Sigma(\tau_1,\tau_2)$. Then at last the path integral is
\begin{equation}
\begin{split}
   \av{Z} = \int D^N\Omega \CD \Sigma \CD G \CD^N h \exp\left(N\frac{J^2 }{2 q} \int G^qd\t^2 - \frac{N}{2} \int \Sigma G d\t^2+ \int \sum_{r\mu} s_{r\mu} h_{r\mu} \right)=\\
   \int \CD \Sigma \CD G  \exp\left(N\frac{J^2 }{2 q} \int G^qd\t^2 - \frac{N}{2} \int \Sigma G d\t^2+ N\log \langle\mathcal P e^{\int \sum_\mu \sigma_\mu h_\mu(\tau) d\t}\rangle_{\Sigma} \right)
\end{split}
   \label{eq:GSigma}
\end{equation}
where $\CD h$ is a Gaussian measure with covariance $\Sigma$ and $\langle\mathcal P e^{\int \sum_\mu \sigma_\mu h_\mu(\tau) d\t}\rangle_{\Sigma}$ denotes a single spin path integral averaged over magnetic fields. For a full writing of the path integral, see appendix \ref{app:PathIntegral}.

The equations of motion (EOM) or saddle-point equations are
\begin{equation}
\begin{split}
    \Sigma = J^2 G^{q-1} \,\,\,\, (\text{$G$ EOM})\\
    G = \langle [\sigma_x(\tau_1) \sigma_x(\tau_2) + \sigma_y(\tau_1) \sigma_y(\tau_2)  + \sigma_z(\tau_1) \sigma_z(\tau_2)] \rangle_{\Sigma} \,\,\,\, (\text{$\Sigma$ EOM})
\end{split}
\label{eq:EOMdiag}
\end{equation}
where $\langle \cdots \rangle_{\Sigma}$ denotes a single-spin correlator in the presence of a random (imaginary) time-dependent field $h(\tau)$ with covariance $\Sigma$. This correlator should be evaluated as an annealed average over $h$, meaning we separately average the numerator and denominator. All the saddles we consider will have time-translation symmetry, meaning $G(\tau_1,\tau_2) = G(\tau_1 - \tau_2)$ and $\Sigma(\tau_1,\tau_2) = \Sigma(\tau_1 - \tau_2)$, unless otherwise noted.

\subsection{Sampling Approach}

Practically speaking, it is still hard to deal analytically with the fluctuating $h$ field. However, we can adopt a sampling approach. We divide $\tau \in [0,\beta)$ up into $N_\tau$ intervals of length $\Delta \tau = \beta / N_\tau$. One sample corresponds to $3 N_\tau$ numbers $h_\mu(j \Delta\tau)$ drawn from a Gaussian distribution with covariance $\delta_{\mu,\mu'}\Sigma(j \Delta \tau, j' \Delta \tau)$. We then construct the single spin path integral
\begin{equation}
    Z_1 = \tr \prod_j e^{\Delta \tau \sigma_\mu h_\mu(j \Delta \tau)}.
\end{equation}
By averaging over $h$, we get $\av{Z_1}_h$. We also need the analog of $Z_1$ but with the operators inserted, $Z_{1,\text{op}}$, and its average over $h$, $\av{Z_{1,\text{op}}}_h$. The correlator $G$ is
\begin{equation}
    G = \frac{\av{Z_{1,\text{op}}}_h}{\av{Z_1}_h} = \langle \sigma_\mu(\tau_1) \sigma_\mu(\tau_2) \rangle_{\Sigma}.
\end{equation}
We can approximate both numerator and denominator via sampling. This is an annealed quantity where we separately average numerator and denominator. 

The basic algorithm to solve the EOM starts with a guess for $G$ and $\Sigma$, uses $\Sigma$ and sampling to estimate the corresponding $G$, updates $G$ and $\Sigma$ using the other equation of motion and some update rule (e.g. a linear combination of old and new), and then repeats until convergence. There are further details in the numerical implementation. For example, the update rule can be changed dynamically to aid convergence. We will present data obtained from scheme below in Sec.~\ref{sec:Numerics}.

\section{Solving the Saddle-Point Equations}
\label{sec:Solutions}
\subsection{Annealed Solution}
\label{subsec:annealed}
We now discuss solutions to the EOM. Before considering more complicated saddles, let us first discuss the saddle which computes the annealed free energy, and dominates the disorder-averaged partition function at large $\beta$. In order to make the annealed free energy proportional to $\beta$ at large $\beta$, this saddle must have the property that $G(\tau)$ and $\Sigma(\tau)$ have a non-vanishing limit as $\tau ,\beta \to \infty$. If we make the so-called static approximation, neglecting the time-dependence of $G$ and $\Sigma$,  then we see that the $G$ EOM is solved by
\begin{equation}
    G=1, \Sigma=J^2.
\end{equation}
At very large $\beta$, these values give the large $\tau$ limit of $G(\tau)$ and $\Sigma(\tau)$. 

To see that the $G=1$ solution also satisfies the $\Sigma$ EOM at large $\beta$, consider the single spin problem with fluctuating field $h$. Since $\Sigma$ is independent of time (which is approximately valid for most of $\tau$ at large $\beta$), the fluctuating field $h$ can be taken to be uniform in imaginary time. A Gaussian distribution with covariance $\Sigma(\tau,\tau') = \text{constant}$ can only generate constant-in-time $h$ fields. 

Thus, we must analyze a single spin in a uniform random field where each component has variance $J^2$ in the limit of large $\beta$. This is straightforward, as $\beta \to \infty$ simply projects onto the ground state of the $h$ field. In this ground state, $G$ will be unity since, after a rotation of the axes to align the $x$-axis with the random field, only $\sigma_x$ has a non-vanishing contribution,
\begin{equation}
    G \approx \frac{ e^{(\beta - \tau)|h|} \langle \sigma_x \rangle e^{\tau |h|} \langle \sigma_x \rangle  }{e^{ \beta |h|}} = \langle \sigma_x \rangle^2 = 1.
\end{equation}

The actual annealed free energy can be obtained via a simple trick. Being dimensionless, $\av{Z}$ depends only on the dimensionless combination $\beta J$. Hence, the thermodynamic identity 
\begin{equation}
    E_a = - \partial_\beta \ln \av{Z}
\end{equation}
can be converted into 
\begin{equation}
    E_a = - J \partial_{\beta J} \ln \av{Z}  = - \frac{J}{\beta} \partial_J \ln \av{Z}.
\end{equation}
Now, dependence on $J$ comes from two sources, the explicit dependence in front of $G^q$ and the implicit dependence via the saddle point solution. However, the derivative of these fields gives no contribution since the EOM is obeyed,
\begin{equation}
    \partial_J \text{action} = \text{explicit term} + \text{EOM}_G \partial_J G + \text{EOM}_\Sigma \partial_J \Sigma.
\end{equation}
Hence, we need only consider the explicit dependence. The result is
\begin{equation}
    E_a = -\frac{ J^2}{q} \int d\tau G_a(\tau)^q,
    \label{eq:Eint}
\end{equation}
where again $G_a(\tau) \to 1$ as $\tau \to \infty$. Note that this equation is true beyond the static limit, and analogous identities relate other saddles with the corresponding energies. 

Since $G_a$ is approximately constant for most of $\tau$, it follows that
\begin{equation}
    E_a = - \frac{J^2}{q} \left( \beta + \cdots \right)
\end{equation}
at large $\beta$. Because the annealed energy diverges to negative infinity as $\beta \to \infty$, it cannot possibly be equal to the quenched energy as the latter is necessarily bounded given the bounded spectrum of the \name{} Hamiltonian. Hence, the annealed free energy, while correctly computing $\av{Z}$, fails to capture the quenched physics at sufficiently large $\beta$.

\subsection{The Asymptotic Value of $G$}

For the annealed solution at large $\beta$, we found $G(\infty)=1$. What other possible values can $G(\infty)$? Here we argue that the only other possible value for a solution of the single-replica EOM is $G(\infty)=0$. Note that other values are possible for solutions of the multi-replica EOM. Even if other asymptotic values $G(\infty)$ were possible for single-replica solutions, none of these could control the quenched free energy since they would all have diverging energy according to \eqref{eq:Eint}.

We make use of another way of expressing the energy,
\begin{equation}E=\frac{NG'(0^+)}{4q},
\label{eq:Ederiv}
\end{equation}
which follows from $\sum_{r\mu}[H,\sigma_{r\mu}]\sigma_{r\mu}=4qH$. Both this formula and \eqref{eq:Eint} compute the energy of a given saddle and agree even when the saddle is not the dominant one. 

We work at large $\beta$ and suppose that $G(\infty)$ is some non-vanishing constant in this limit. From \eqref{eq:Eint} we learn that $E \sim - J^2 \beta G(\infty)^q$. From \eqref{eq:Ederiv} we learn that $G(\tau)$ must vary rapidly near $\tau=0$ in order to generate such a large $E$, $G'(0)\sim - J^2 \beta G(\infty)^q$. At large $\beta$ and fixed $G(\infty)$, we therefore expect $G$ to be close to its late time value after a very short time interval of size $1/(\beta J^2 G(\infty)^q)$.

This rapid approach to the late time constant value of $G$ implies a similar rapid approach of $\Sigma$ to its late time value. Since $\Sigma -\Sigma(\infty)$ is rapidly decaying, we again appeal to the static approximation with the expectation that it will be quantitatively accurate at large $\beta$. The only possible solution with non-vanishing $G(\infty)$ thus has $G(\infty)=1$. This is the annealed solution we already discussed. The only other possibility is $G(\infty)=0$, for which the static approximation must fail.

\subsection{Non-Annealed Solutions Can't Decay Exponentially}
\label{subsec:noExp}
In this subsection, we argue that the equations of motion don't have any rapidly decaying solutions apart from the annealed solution. For the next two subsections, \ref{subsec:noExp} and \ref{subsec:powerlaw}, we assume that $G(\infty)=0$ at large $\beta$. The impossibility of a rapidly decaying which could control the quenched free energy means that \name{} cannot have a gapped ``paramagnetic'' phase.

As a warmup, we will first show that the model cannot have white noise for which $\Sigma(\t_1,\t_2)\propto \delta(\t_1-\t_2$). We start with the $\Sigma$ equation, written as
\begin{equation}
    G(\t_1,\t_2) = \frac{\av{Z_{1,\text{op}}}_h}{\av{Z_1}_h}.
\end{equation}
The denominator can be written as 
\begin{equation}
    \av{Z_1}_h=\tr \int \mathcal D h_\mu \mathcal P \exp \left(\int_0^\beta h_\mu(t)\sigma_\mu d\t\right),
    \label{eq:denomG}
\end{equation}
while the numerator is
\begin{equation}
\begin{split}
    \av{Z_{1,\text{op}}}_h=3\tr \int \mathcal D h_\mu(0<\t\leq \t_1)\mathcal D h_\mu(\t_1<\t\leq \t_2)\mathcal D h_\mu(\t_2<\t\leq\beta)\\ 
    \mathcal P \exp \left(\int_0^{\t_1} h_\mu(\t)\sigma_\mu d\t\right)\exp \left(\int_{\t_1}^{\t_2} \tilde h_\mu(\t)\sigma_\mu d\t\right)\exp \left(\int_{\t_2}^{\beta} h_\mu(\t)\sigma_\mu d\t\right),
    \label{eq:numG}
\end{split}
\end{equation}
with $\tilde h$ being $h$ rotated 180 degrees around a given axis. If $\Sigma$ is a delta function, then $\int \mathcal D h=\int \mathcal D \tilde h$, so after relabeling our variables the numerator in equation \eqref{eq:numG} equals three times the denominator in equation \eqref{eq:denomG}, we find $G(\t_1,\t_2)=3$. This is obviously incompatible with a delta function $\Sigma$.

If $\Sigma$ and $G$ decay exponentially, then $h$s at times separated by more than a few decay times are essentially independent. The boundary effects can change the integral by at most a multiplicative constant, so we have 
\begin{equation}
    \av{Z_{1,\text{op}}}_h=\alpha \av{Z_1}_h
\end{equation}
 for some constant $\alpha$. Hence, starting from our assumption that $G$ decays exponentially, we've shown that it plateaus as some fixed $\alpha$, thus reaching a contradiction and showing that $G$ cannot decay exponentially.
 
\subsection{A Power-Law Saddle}
\label{subsec:powerlaw}
We have so far found an annealed solution and just ruled out a solution which decays exponentially to zero. The remaining possibility is a saddle with power-law decay.

To explain why such a saddle in fact exists, it is useful analyze the path integral as expressed in equation \eqref{eq:GSigma}. From this point of view, the integral over $\Omega_\mu(\tau)$ computes the statistical partition function of a classical XYZ chain with couplings given by $\Sigma(\tau_1 -\tau_2)$. Here the chain direction is imaginary time and the Green function $G$ is obtained as the statistical correlation in this XYZ model. The fact that $\Sigma=J^2G^{q-1}$ imposes the non-trivial constraint that the XYZ couplings ($\Sigma$) are related to the spin correlations ($G$). We also note that the measure $\CD \Omega$ contains strong nearest couplings which produce perfect ordering in the absence of $\Sigma$, but this is not a self-consistent solution except at $\beta J =0$.

Let us denote the power-law couplings as 
\begin{equation}
    \Sigma(\tau) \sim \frac{\Sigma_0}{|\tau|^{1+s}}
\end{equation}
for some to-be-determined parameter $s$. This implies a corresponding power-law in $G$ thanks to the $G$ EOM. To determine $s$ we must obtain another expression for $G$ by solving the $\Sigma$ EOM, i.e. by solving the statistical spin chain problem. We work in the large $\beta$ limit for simplicity. Remember that $\beta$ corresponds to the length of the chain, and the role of inverse temperature in the stat mech picture of the chain is played by the magnitude of $\Sigma_0$.

As a function of $\Sigma_0$, this model has a phase transition from an ordered state at large $\Sigma_0$ to a disordered state at small $\Sigma_0$. The ordered state boasts long-range order in $G$ and so corresponds to a non-vanishing value of $G$ at large $\tau$. However, this does not yield a self-consistent solution of the $G$ EOM. At small $\Sigma_0$, the $G$ correlator decays in the same way as $\Sigma$, like $\tau^{-(1+s)}$ (note the sign of $s$), which is also not a self-consistent solution. The remaining possibility is that $\Sigma_0$ sits right at a phase transition between the ordered phase and the disordered phase. 

Going back to work of Dyson~\cite{dyson_ising_lr}, this ``long-range fixed point'' has been analyzed extensively (including in more than one dimension)~\cite{PhysRevLett.29.917,PhysRevB.8.281,Honkonen_1989,Angelini_2014,Paulos_2016,Behan_2017}. The most important feature for our purposes is that, at the critical point, for $\Sigma$ decaying as $\tau^{-(1+s)}$, the corresponding correlations decay as
\begin{equation}
    G(\tau) \sim \frac{1}{\tau^{1-s}}.
\end{equation}
This is why we introduced the $s$ parameter above. Remarkably, the long-range character of the fixed point means this relation, which is what we would have for a Gaussian field theory, is valid even with interactions.
For $s \leq 0$, the couplings decay too slowly and the system is always ordered. For $s \geq 1$, the couplings decay too rapidly and one recovers the short-range physics. There is no phase transition for the short-range XYZ model in 1d. So the interesting range of $s$ is 
\begin{equation}
    0 < s < 1.
\end{equation}
It is also known that for $s <1/2$, the fixed point is Gaussian, while for $s>1/2$, the Gaussian fixed point is unstable and flows to an interacting long-range fixed point.

In our case, the $G$ EOM requires 
\begin{equation}
    (1-s)(q-1) = 1+s,
\end{equation}
which yields
\begin{equation}
    s = 1 - \frac{2}{q}.
\end{equation}
The corresponding scalings are 
\begin{equation}
    G \sim \frac{1}{\tau^{2/q}}
\end{equation}
and 
\begin{equation}
    \Sigma \sim \frac{1}{\tau^{2(1 - 1/q)}},
\end{equation}
which are precisely the SYK correlations in the ground state. Of course, this is essentially a result of dimensional analysis once we know the Gaussian result for the power-law exponent is robust.

We refer to this saddle as SYK-like. If it is the saddle selected by minimax, then the physics of \name{} at low temperature will share at least some similarities with the SYK. In particular, it will feature an emergent conformal symmetry, as the long-range fixed points considered above are known to enjoy an enhanced conformal symmetry which includes the scaling symmetry~\cite{Paulos_2016}.

\subsection{No $q=2$ SYK-like Saddle}

It is interesting to study the SYK-like saddle as a function of $q$. In particular, for $q=2$ we see that $s=0$. This corresponds to couplings that decay as $1/\tau$, but such a slow decay actually results in a chain that is always in an ordered phase in the thermodynamic limit. Hence, we immediately learn that there actually is no SYK-like saddle for $q=2$. This is good because the physics of the $q=2$ SYK model is indeed very different. For the \name{} model, it seems the only option for $q=2$ is glassy physics.

For $q=3$, we find $s=1/3$ which is in the Gaussian regime. Hence, we can replace the XYZ chain by a Gaussian field theory and obtain explicit Schwinger-Dyson like equations for $G$ in terms of $\Sigma$. For $q=4$, the value of $s$ is $1/2$, which is right on the boundary between the Gaussian and interacting regimes. Interestingly, once we have $q=5$ or greater, the fixed points are all interacting. Hence, there is a significant difference between the physics of the SYK-like saddle for $q \geq 5$ and $q=3,4$. 

\subsection{Summary of Solutions}

We first discussed the annealed solution and then searched for solutions which could control the quenched free energy. We argued that the only possible values of $G(\infty)$ were the annealed value and zero. Moreover, any replica diagonal saddle which controls the quenched free energy must have $G(\infty)=0$ (otherwise the energy \eqref{eq:Eint} would diverge with $\beta$). Among replica diagonal solutions with $G(\infty)=0$, we showed that an exponentially decaying solution was not possible. However, for $q>2$, there is an interesting solution with power-law correlations. 

Hence, we have the following structure for the quenched free energy. For $q=2$, the only available solution is a replica non-diagonal one (which we did not construct). This strongly suggests that $q=2$ is generically glassy, and this is confirmed by evidence from exact diagonalization. This conclusion is broadly consistent with the presence of glassy physics in many quantum spin models with random $2$-local couplings, e.g.~\cite{SY}.

For $q>2$, we do have the SYK-like saddle, but we may also have replica non-diagonal saddles. The correct physics is determined by minimax. We have not yet carried out the full minimax procedure, so here we rely on evidence from exact diagonalization (see Sec.~\ref{sec:ed}) to determine the fate of the system for $q>2$. 

To substantiate the general arguments in this section, we first discuss the single-replica solutions in more detail using an analytical large $q$ approach (Section~\ref{sec:Liouville}) and a numerical approach (Section~\ref{sec:Numerics}). Then we return to the exact diagonalization analysis in Section~\ref{sec:ed}.

\section{Large $q$ Limit and the Liouville equation}
\label{sec:Liouville}

To gain further insight into the problem (and because it will be useful for numerically solving the EOM), we can study the large $q$ limit \cite{Maldacena_2016,Tarnopolsky}. We start with the equations of motion
\begin{equation}
    \Sigma(\t_1,\t_2) = J^2 G^{q-1}(\t_1,\t_2)
\end{equation}
\begin{equation}
    G(\t_1,\t_2) = \langle \sigma_\mu(\t_1) \sigma_\mu(\t_2) \rangle_\Sigma
\end{equation}
and we make the ansatz
\begin{equation}
\begin{split}
    G = 3 \left( 1 + \frac{g}{q} + \cdots\right),\\
    \Sigma(\tau) = 3^{q-1} J^2 e^{g(\tau)} + \cdots.
\end{split}
\end{equation}
To close the EOM, we need to compute $G$ in terms of $\Sigma$.

This can be done conveniently starting with a formula for the derivatives of $G$,
\begin{equation}
    \partial_{\t_1}\partial_{\t_2} G = \frac{\int \CD h \tr(\mathcal{P}\left\{e^{\int h\cdot \sigma} [h_\nu(\tau_1) \sigma_\nu(\tau_1) ,\sigma_\mu(\tau_1)][h_\rho(\tau_2) \sigma_\rho(\tau_2), \sigma_\mu(\tau_2)]\right\})}{\int \CD h \tr(\mathcal{P}\left\{e^{\int h\cdot \sigma} \right\})}.
    \label{eq:G_dd}
\end{equation}
In the perturbative limit where $G$ is close to $3$, we expand \eqref{eq:G_dd} perturbatively in $\Sigma$. Since the numerator already contains a factor $\Sigma$ that arises from the $h_\nu(\t_1) h_\rho(\t_2)$ factors, we can neglect the fields in the exponents to obtain
\begin{equation}
     \partial_{\t_1}\partial_{\t_2} G \approx \frac{ \Sigma(\tau_1,\tau_2) \tr( [\sigma_{\nu},\sigma_{\mu}] [\sigma_{\nu},\sigma_{\mu}]) }{\tr(1)} = - 24 \Sigma(\tau_1,\tau_2).
\end{equation}
On the other hand, we have
\begin{equation}
    \partial_{\t_1}\partial_{\t_2} G \approx  \frac{3}{q} \partial_{\t_1}\partial_{\t_2} g 
\end{equation}
from the large $q$ ansatz, so we can now close the EOM,
\begin{equation}
    \partial_{\t_1}\partial_{\t_2} g = - 8 q 3^{q-1} J^2 e^{g}.
\end{equation}

We introduce the new coupling
\begin{equation}
    \mathcal{J}^2 = 4q 3^{q-1} J^2.
    \label{eq:calJ}
\end{equation}
and make the further ansatz $g(\t_1,\t_2) = g(\t_1-\t_2)$, to finally obtain
\begin{equation}
    \partial_\t^2 g = 2 \mathcal{J}^2 e^g,
\end{equation}
which is the Liouville equation. The standard solution is
\begin{equation}
    e^g = \left[ \frac{\cos \frac{\pi v}{2}}{\cos \left[ \pi v \left(\frac{1}{2} - \frac{\tau}{\beta}\right)\right]}\right]^2
\end{equation}
\begin{equation}
    \beta \mathcal{J} = \frac{\pi v}{\cos \frac{\pi v}{2}}.
\end{equation}
Note that this has the right boundary values, with $g$ vanishing at $\tau=0,\beta$.

Using \eqref{eq:Ederiv} for the energy, we can immediately obtain the thermodynamics from the derivative of $g$ at $\tau=0^+$. As an example, in the $\beta \to \infty$ limit, the standard solution limits to
\begin{equation}
    e^g = \frac{1}{(\mathcal{J} \tau +1)^2}.
\end{equation}
Near $\tau=0$ we thus have $g = - 2 \mathcal{J} \t + \cdots$, and so
\begin{equation}
    - \frac{E}{N} = - \frac{G'(0^+)}{4q} = \frac{3 \mathcal{J}}{2 q^2} .
\end{equation}
We also get a large-$q$ expression for the ground state entropy analogous to the SYK result:
\begin{equation}
    S_0/N\approx \log 2-\frac {3\pi^2}{8q^2}
\end{equation}

It is also useful to perform an approximate resummation of the large $q$ expansion in which we replace
\begin{equation}
   G = 3\left( 1 + \frac{g}{q} + \cdots \right) \to 3 e^{g/q}.
\end{equation}
For example, at large $\beta$ this gives 
\begin{equation}
    G \to \frac{3}{(\mathcal{J} \tau +1)^{2/q}},
\end{equation}
which clearly displays a scaling behavior with $G \sim \tau^{-2/q}$. This resummation can be understood as arising from an estimate of \eqref{eq:G_dd},
\begin{equation}
    \partial_{\t_1} \partial_{\t_2} G \approx - 8 \Sigma(\t_1,\t_2) G(\t_1,\t_2).
\end{equation}
Using the short-time form of $G$ gives the previous result of $- 24 \Sigma$, but more generally we can make the ansatz $G = 3 e^{g/q}$ and compute
\begin{equation}
   \partial_{\t_1} \partial_{\t_2} G = \left( \frac{1}{q} \partial_{\t_1} \partial_{\t_2} g + \frac{1}{q^2} \partial_{\t_1} g \partial_{\t_2} g  \right) G.
\end{equation}
At large $q$, we neglect the second term and again obtain the Liouville equation.

\section{Numerical Solution of the Equations of Motion}
\label{sec:Numerics}

Here we discuss approximate numerical solutions of the EOM obtained from an iterative approach. A similar approach was used many years ago in the $q=2$ case~\cite{PhysRevLett.80.389}. The precise numerical details are described in App.~\ref{app:num_methods}. We considered two kinds of initializations for the iterative solver. The first is the $J=0$ solution, $G(\tau)=3$, which often converges to the annealed saddle at sufficiently large $\beta$. For the second, we found it very useful to initialize with the analytic large-$q$ form obtained above. Specifically, we exponentiate the standard solution to give
\begin{equation}
    G_{\text{large}} =  3 \left[ \frac{\cos \frac{\pi v}{2}}{\cos \left[ \pi v \left(\frac{1}{2} - \frac{\tau}{\beta}\right)\right]}\right]^{2/q}.
\end{equation}
This second type of initialization typically converges to something consistent with the interesting quenched saddle discussed above, even if the initial guess is not particularly close, i.e. $q$ is not large. 

The basic structure of the solver is as follows. The imaginary time circle is discretized into $N_\tau$ units of length $\Delta \tau = \beta/N_\tau$ and we assign a value of $G$ for each time point $\tau_\ell =  \ell \Delta \tau$. Similarly, $\Sigma$ is an $N_\tau \times N_\tau$ matrix. We enforce translation symmetry in $G$ and $\Sigma$ and the $\tau \leftrightarrow \beta - \tau$ symmetry in $G$. The method works by proposing an update, $G_{\text{prop}}, \Sigma_{\text{prop}}$, based on the current values, $G_{\text{curr}},\Sigma_{\text{curr}}$. The proposed new $\Sigma$ is simply $\Sigma_{\text{prop}} = J^2 G_{\text{curr}}^{q-1}$. The proposed new $G$ is obtained by sampling the single-spin problem with magnetic fields of covariance $\Sigma_{\text{curr}}$. Then $G$ is updated to some linear combination of $G_{\text{curr}}$ and $G_{\text{prop}}$ and similarly for $\Sigma$. To assess the degree of convergence, we monitor an error given by
\begin{equation}
    \varepsilon = \sqrt{\frac{1}{N_\tau} \sum_\ell (G_{\text{curr}}(\tau_\ell) - G_{\text{prop}}(\tau_\ell))^2 } + \sqrt{\frac{1}{N_\tau} \sum_\ell (G_{\text{curr}}(\tau_\ell)^{q-1} - G_{\text{prop}}(\tau_\ell)^{q-1})^2 }.
\end{equation}
Note that this error is intensive in $N_\tau$, so it represents an error per site. When it is small, then the proposed update is close to the current value on a site-by-site basis. Practically speaking, we can never make this error smaller than our sampling error.

In addition to the large $q$ analytics, we also compare with the SYK conformal solution,
\begin{equation}
    G_{\text{conf}} = b \left( \frac{\pi}{\beta \sin \frac{\pi \tau}{\beta}} \right)^{2/q},
\end{equation}
with $J^2 b^q = \frac{\left(\frac{1}{2} - \frac{1}{q}\right) \tan \frac{\pi}{q}}{\pi}$. A priori, even if the \name{} model has a power-law solution, it would not have to be exactly equal to the SYK solution, but in practice we find that this is the case.

\begin{figure}
    \centering
    \includegraphics[width=.8\textwidth]{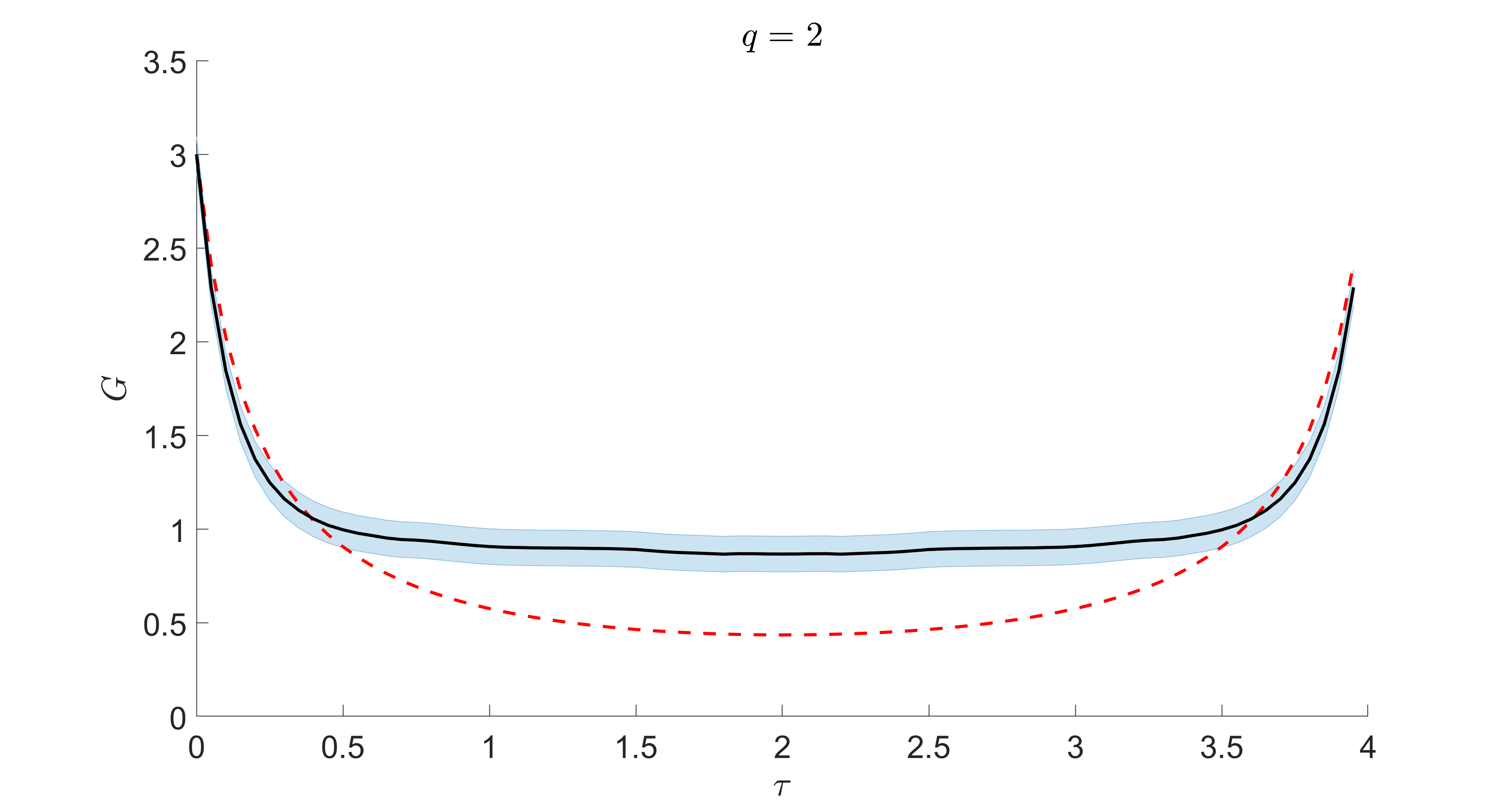}
    \caption{Solution to the EOM with $q=2$, $\beta=4$, $J=1$, $N_\tau=80$, and $N_s=1000$ samples. The red dashed line is the initialization (large $q$), the black line shows the final converged result, and the blue shaded region gives $3/\sqrt{N_s}$ around the black line. The result is consistent with the annealed saddle in which $G$ approaches unity at long time. }
    \label{fig:q2_eom_ann}
\end{figure}

\begin{figure}
    \centering
    \includegraphics[width=.8\textwidth]{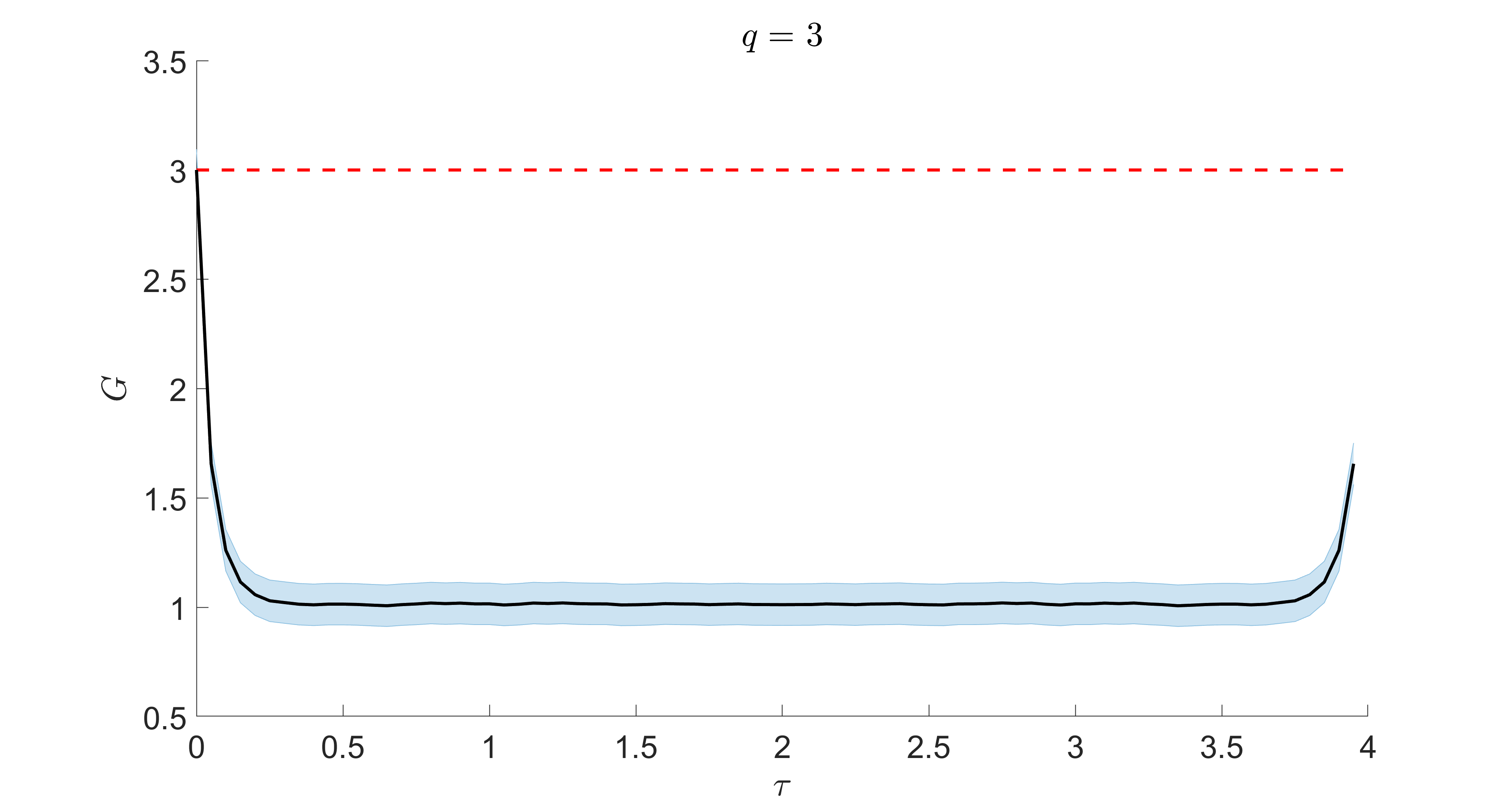}
    \caption{Solution to the EOM with $q=3$, $\beta=4$, $J=1$, $N_\tau=80$, and $N_s=1000$ samples. The red dashed line is the initialization ($J=0$), the black line shows the final converged result, and the blue shaded region gives $3/\sqrt{N_s}$ around the black line. The result is consistent with the annealed saddle in which $G$ approaches unity at long time. }
    \label{fig:q3_eom_ann}
\end{figure}

Figure~\ref{fig:q2_eom_ann} and Figure~\ref{fig:q3_eom_ann} show annealed-type solution for $q=2$ and $q=3$, respectively. These solutions can be obtained with modest computational effort. We also see that the static approximation looks good, and we expect it will only improve with increasing $\beta$. For $q=2$, this is the only type of solution we have been able to find.

\begin{figure}
    \centering
    \includegraphics[width=.8\textwidth]{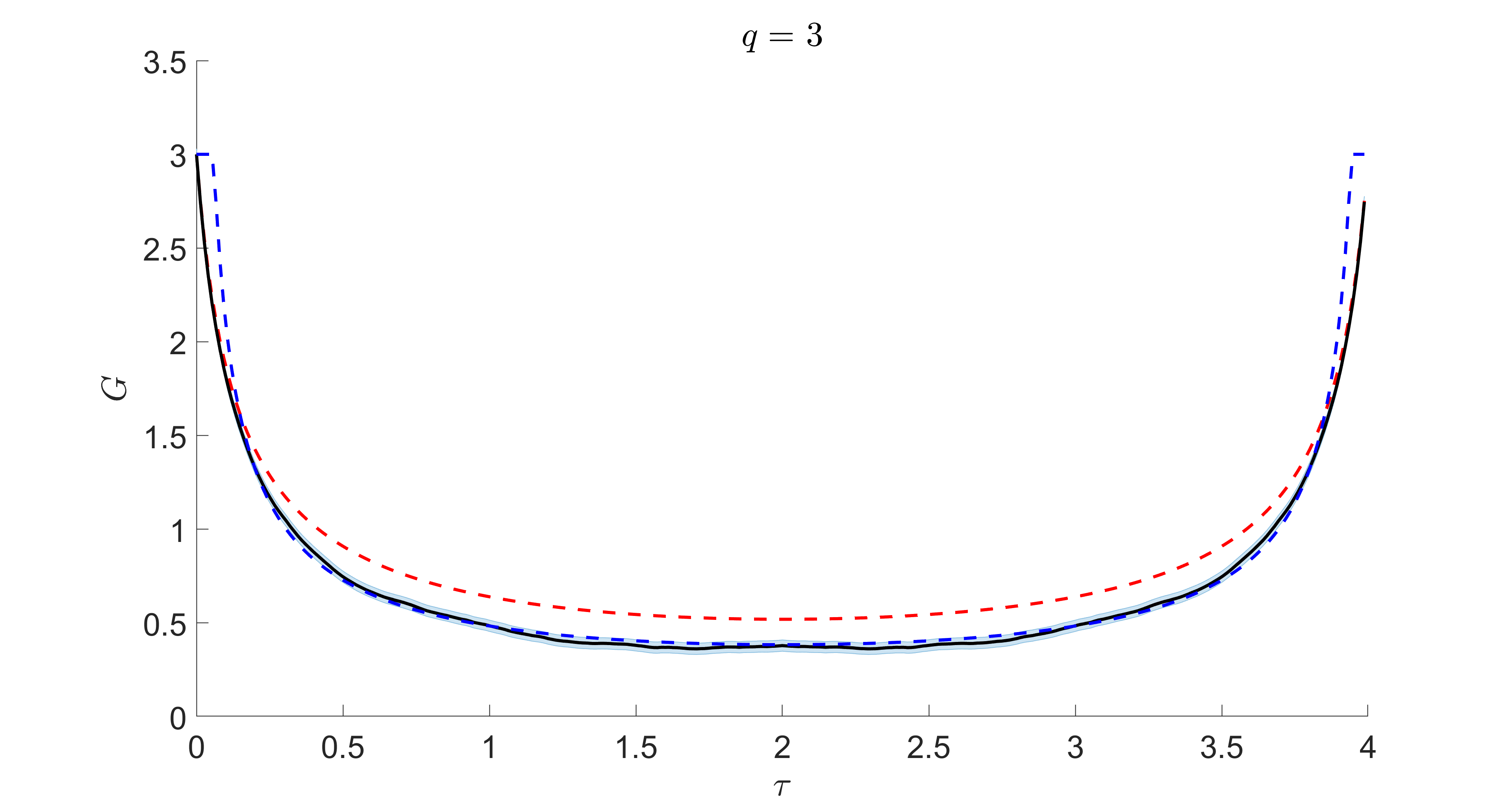}
    \caption{Solution to the EOM with $q=3$, $\beta=4$, $J=1$, $N_\tau=300$, and $N_s=10000$ samples. Compared to the previous figures, we have now dramatically increased the number of time points and the number of samples. The red dashed line is the initialization (large $q$), the black line shows the final converged result, and the blue shaded region gives $3/\sqrt{N_s}$ around the black line. The blue dashed line is the SYK conformal result (capped at $3$). We see quite good agreement with the large $q$ result at short times and the SYK result at long times. Note how the solution goes well below the annealed value at large $\tau$. }
    \label{fig:q3_eom_syk}
\end{figure}

For $q=3$, one can find other solution types using a large $q$ initialization. Such a solution is shown in Figure~\ref{fig:q3_eom_syk} where we also compare it with the SYK conformal result. The conformal result is an excellent approximation except very close to $\tau=0,\beta$ (where the large $q$ ansatz is closer). We also find that the computational resources need to be significantly increased to find such solutions reliably, including more time points and more samples to estimate $G$ from $\Sigma$. Readers familiar with SYK may be surprised that $\beta=4$ seems already solidly in the conformal regime, but this can be understood because the energy scales of \name{} are enhanced by a large factor arising from the different normalizations of the operators and the couplings.

\begin{figure}
    \centering
    \includegraphics[width=.8\textwidth]{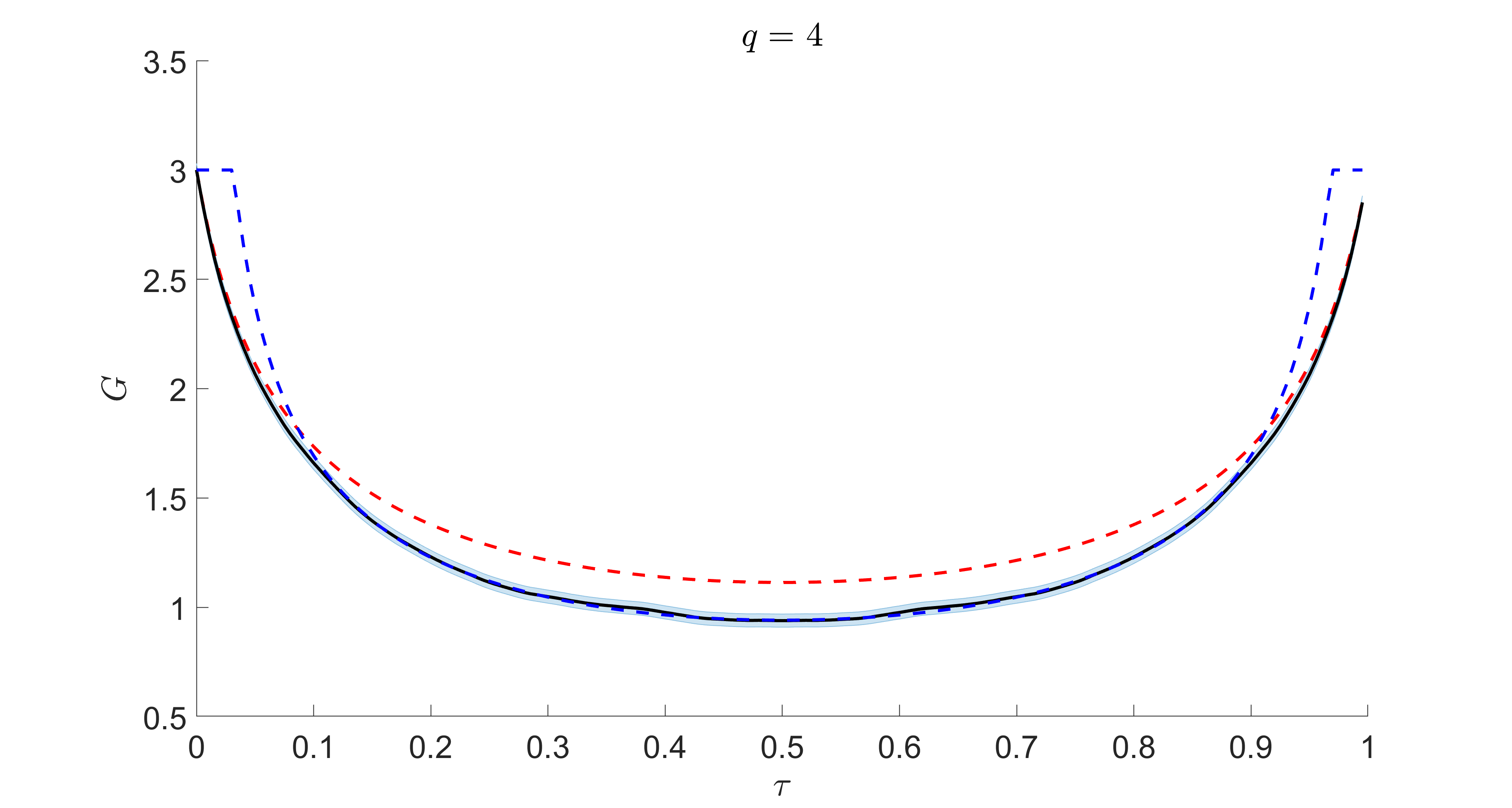}
    \caption{Solution to the EOM with $q=4$, $\beta=1$, $J=1$, $N_\tau=200$, and $N_s=10000$ samples. The red dashed line is the initialization (large $q$), the black line shows the final converged result, and the blue shaded region gives $3/\sqrt{N_s}$ around the black line. The blue dashed line is the SYK conformal result (capped at $3$). This result is roughly consistent with the large $q$ form for all $\tau$.  We again see good agreement with the large $q$ result at short times and the SYK result at long times. }
    \label{fig:q4_eom_syk}
\end{figure}

We can find similar solutions for $q=4$ as shown in Figure~\ref{fig:q4_eom_syk}. Here we set $\beta=1$ instead of $\beta=4$ in the $q=3$ and still find excellent agreement with the conformal ansatz. This is again demonstrating the very rapid rise in energy scale with increasing $q$, c.f. \eqref{eq:calJ}. We can presumably find such a solution for any $q$, although the computational cost seems to continue to increase with increasing $q$, at least with our current approach.

\begin{figure}
    \centering
    \includegraphics[width=.8\textwidth]{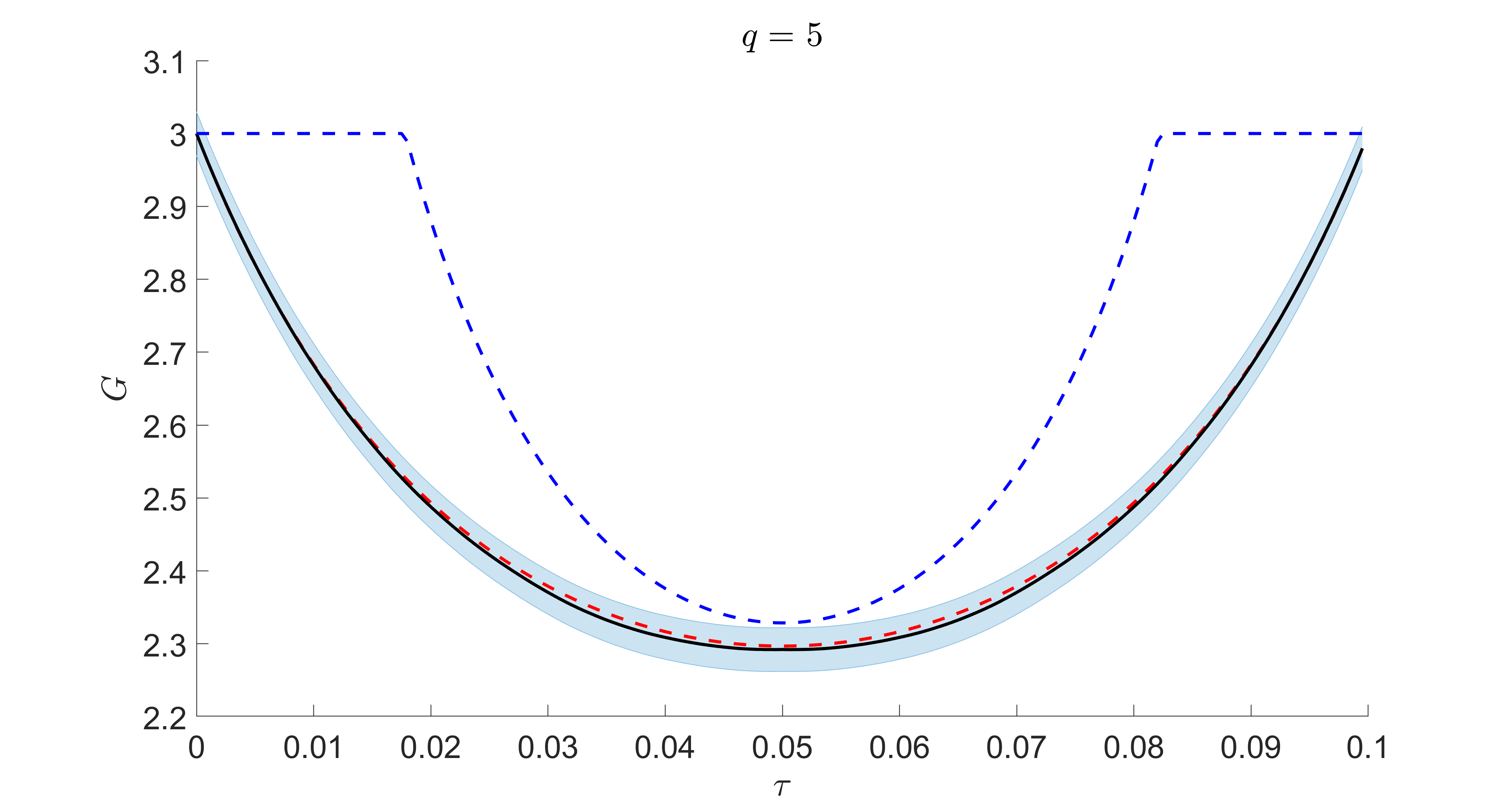}
    \caption{Solution to the EOM with $q=5$, $\beta=.1$, $J=1$, $N_\tau=200$, and $N_s=10000$ samples. The red dashed line is the initialization (large $q$), the black line shows the final converged result, and the blue shaded region gives $3/\sqrt{N_s}$ around the black line. The blue dashed line is the SYK conformal result (capped at $3$). The blue dashed line is the SYK conformal result (capped at $3$). We have significantly reduced $\beta$ relative to the previous figures to demonstrate the good agreement with the large $q$ result.}
    \label{fig:q5_eom_largeq}
\end{figure}

Finally, we show an example with $q=5$ and even smaller $\beta=.1$ in Figure~\ref{fig:q5_eom_largeq}. For this choice of parameters, we are not in the conformal regime but the large $q$ ansatz gives a good account of the solution.

\pagebreak

\section{Finite Size Numerics on Sparse \name{}}
\label{sec:ed}

In this section we define and study a sparse version of \name{} using sparse matrix techniques \cite{xu2020sparse,garciaSparse,Tezuka_2023}. We consider a simple variant of the sparse model defined by randomly pruning all but $\kappa N$ of the ${N\choose q} 3^q$ terms from the Hamiltonian. The remaining couplings are Gaussian with a renormalized variance adjusted so that $\av{\tr(H^2)}$ has the same value in the dense and sparse models. For the dense model, the second moment of $H$ is
\begin{equation}
    \av{\tr(H^2)/2^n} = \av{J_{r\mu\cdots}^2} \binom{N}{q} 3^q = N J^2 \frac{ 3^q}{q}.
\end{equation}
The couplings in the sparse model have variance $\tilde{J}^2$ obeying $\kappa N \tilde{J}^2 = N J^2 \frac{ 3^q}{q}$, which gives
\begin{equation}
   \tilde{J}^2 = J^2 \frac{3^q}{\kappa q}.
\end{equation}
We have observed little variation in observables as $\kappa$ is varied. In the first set of numerical data discussed below, we set $\kappa=N$, which corresponds to a model that is still relatively sparse but which at large $N$ is likely dense enough largely replicate the physics of the dense model, e.g. $1/\kappa$ corrections to intensive quantities should vanish at large $N$.

We consider three observables in the sparse model. 1) The ground state Edwards-Anderson order parameter \cite{ParisiEA,SompolinskyEA,sommers1982}: $\q=\frac 1N \sum_{r\mu} \left(\bra{\psi}\sigma_{r\mu}\ket \psi\right)^2$ for the many-body ground state $\psi$. This order parameter serves as a diagnostic for glassy behavior, being nonzero for glasses but zero for spin liquids or gapped paramagnets. 2) The gap between the ground state and the first excited state. This would have three different behaviors for the three possible phases. For a gapped paramagnet, we would most likely see a sharp peak. For a spin glass, we would see a continuous distribution down to zero \cite{winer_2022,barney_2023}. For an SYK-like system with both level repulsion \cite{haake2010quantum,berry1977level,doi:10.1063/1.1703775,wigner1959group,mehta2004random,PhysRevLett.52.1,garcia_syk_finiteN,cotler_bh_rmt} and an extensive ground state entropy, one would expect a smooth distribution going to zero as the gap goes to zero. 3) The ground state energy. This is a fairly self-explanatory quantity. It provides a sanity check on our overall understanding of the system and can be used to compare with large $q$ estimates.

\begin{figure}
    \centering
    \includegraphics[width=.8\textwidth]{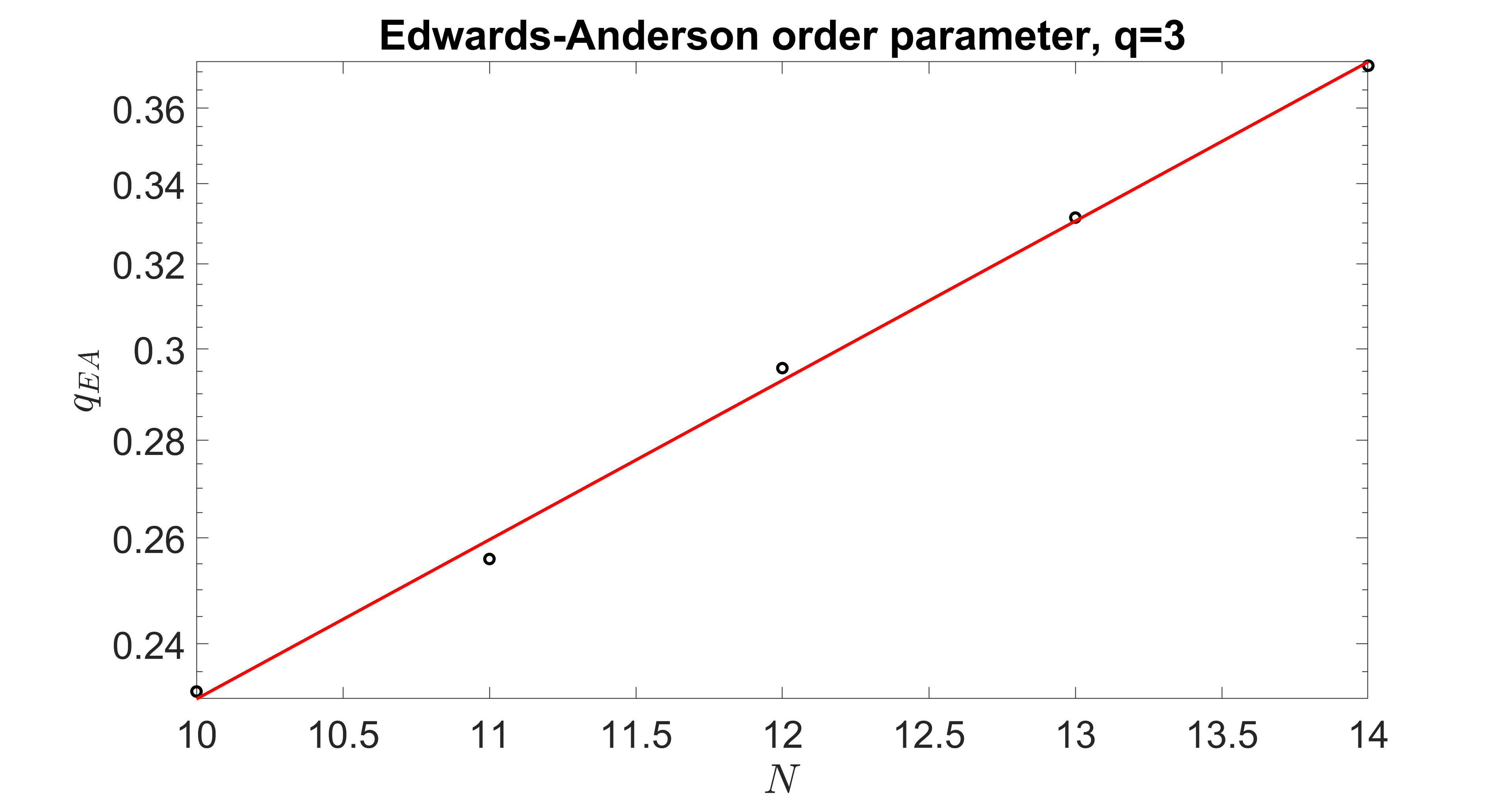}
    \caption{The Edwards-Anderson order parameter in the ground state  (log scale) for $q=3$ as a function of system size. Each point is averaged over $2000$ samples. The rapid growth with system size strongly indicates glassiness. The red line is a linear fit to $\log \q$.}
    \label{fig:q3Qea}
\end{figure}
\begin{figure}
    \centering
    \includegraphics[width=.8\textwidth]{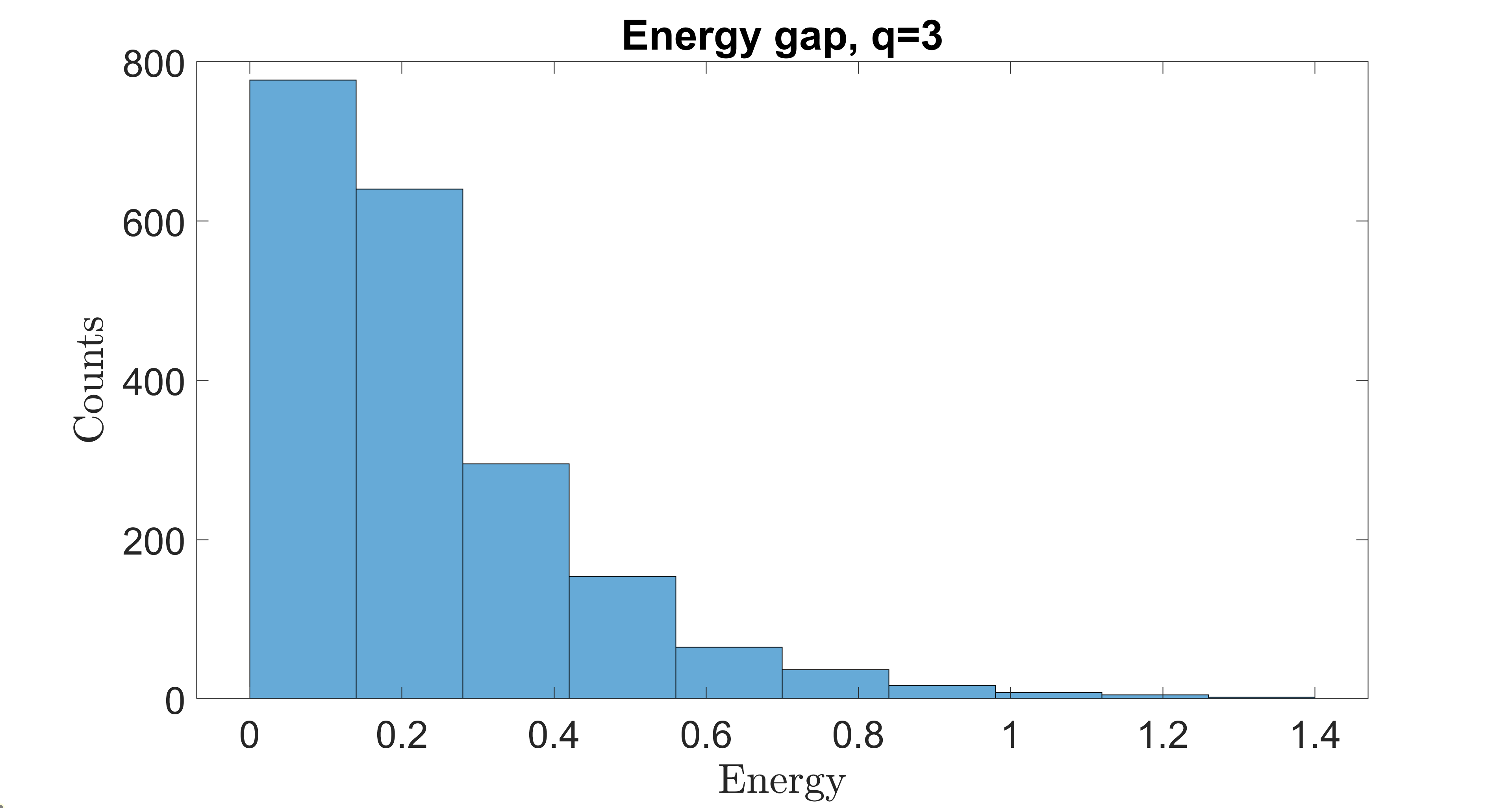}
    \caption{A histogram of $2000$ samples of the finite size energy gap between the ground state and first excited state for $q=3$ and $n=14$. The exponential distribution of the gap and the corresponding lack of level repulsion are consistent with the two lowest lying states being quantum TAP states.}
    \label{fig:level3}
\end{figure}

We discuss $q=3$, $q=4$, and $q=5$ in turn, starting with $q=3$. The EA order parameter is shown in Figure~\ref{fig:q3Qea} where we see an exponential rise as a function of system size. This indicates that the $q=3$ ground state is a glass. This conclusion is supported by the gap histogram in Figure~\ref{fig:level3}. This is exactly what we would expect if the ground state and the first excited state are distinct TAP states with a macroscopic barrier to tunneling between them.

\begin{figure}
    \centering
    \includegraphics[width=.8\textwidth]{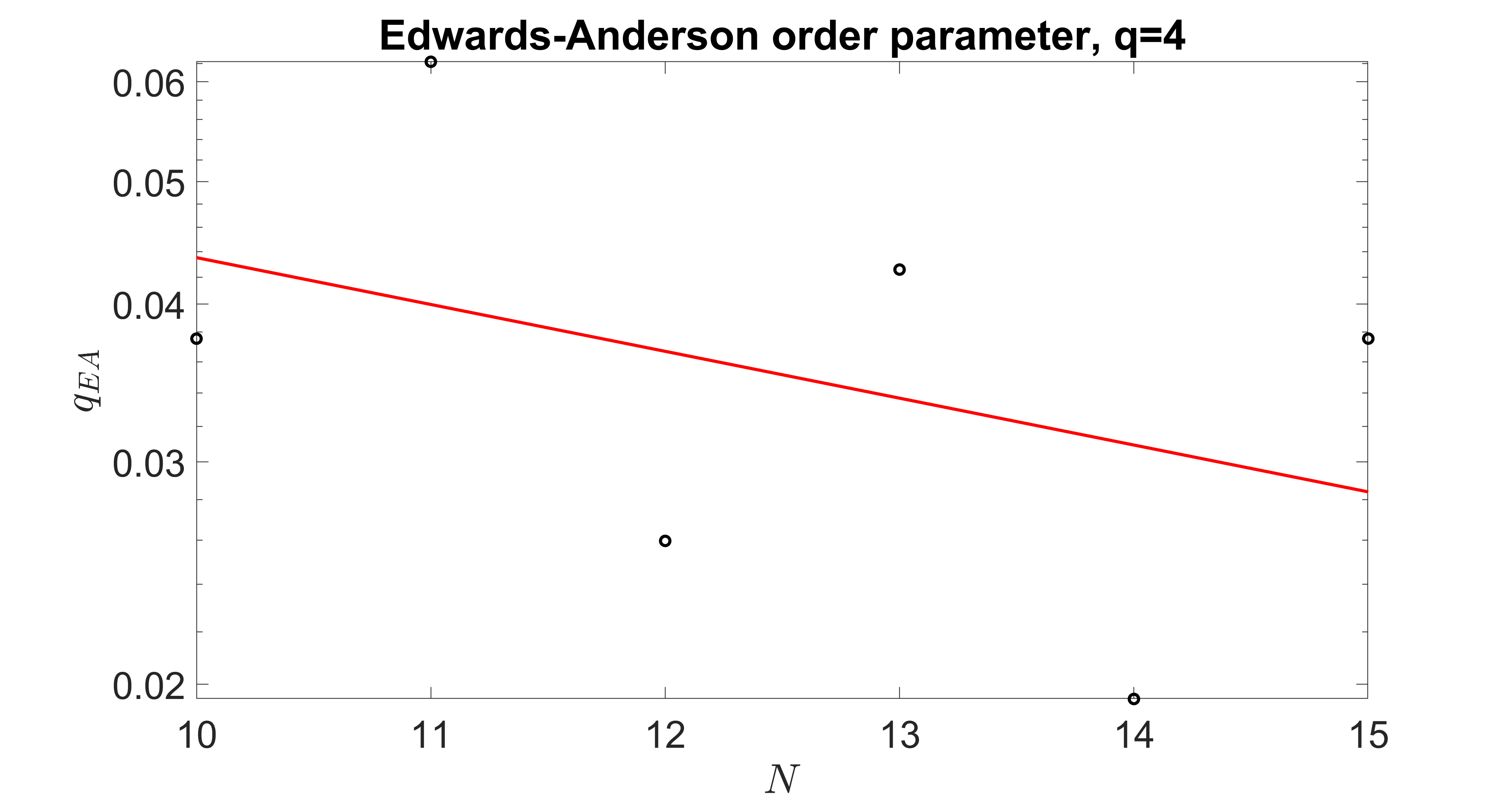}
    \caption{The Edwards-Anderson order parameter in the ground state (log scale) for $q=4$ as a function of system size. Each point is averaged over $500$ samples. We slow decrease with system size which is consistent with the absence of glassiness. We also see a very clear even-odd effect consistent with the anti-unitary time-reversal symmetry. The red line is a linear fit to $\log \q$.}
    \label{fig:q4Qea}
\end{figure}

\begin{figure}
    \centering
    \includegraphics[width=.8\textwidth]{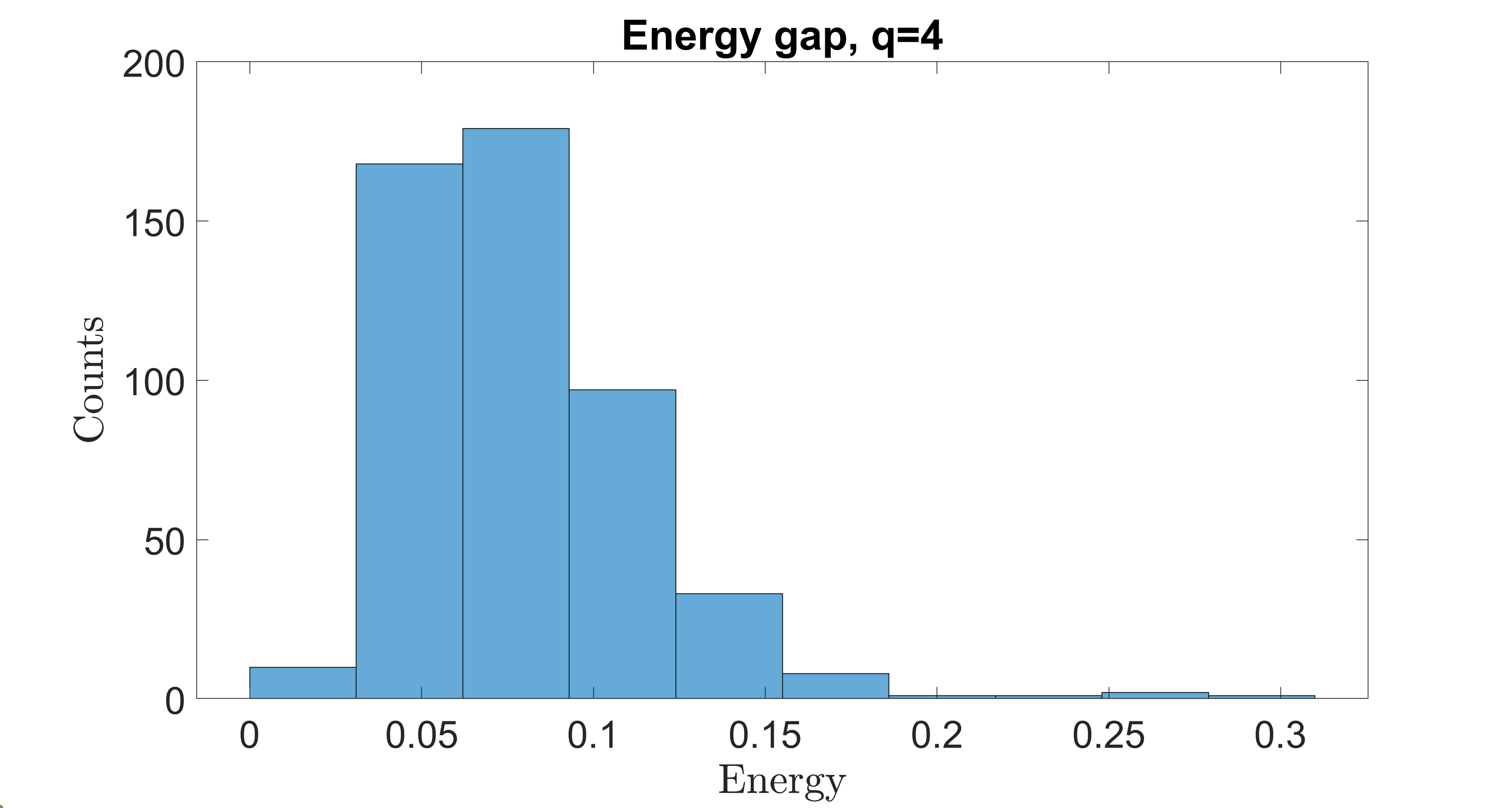}
    \caption{A histogram of $500$ samples of the finite size energy gap between the ground state and first excited state for $q=4$ and $n=15$. The clear level repulsion is consistent with a non-glassy low energy state.}
    \label{fig:level4}
\end{figure}

\begin{figure}
    \centering
    \includegraphics[width=.8\textwidth]{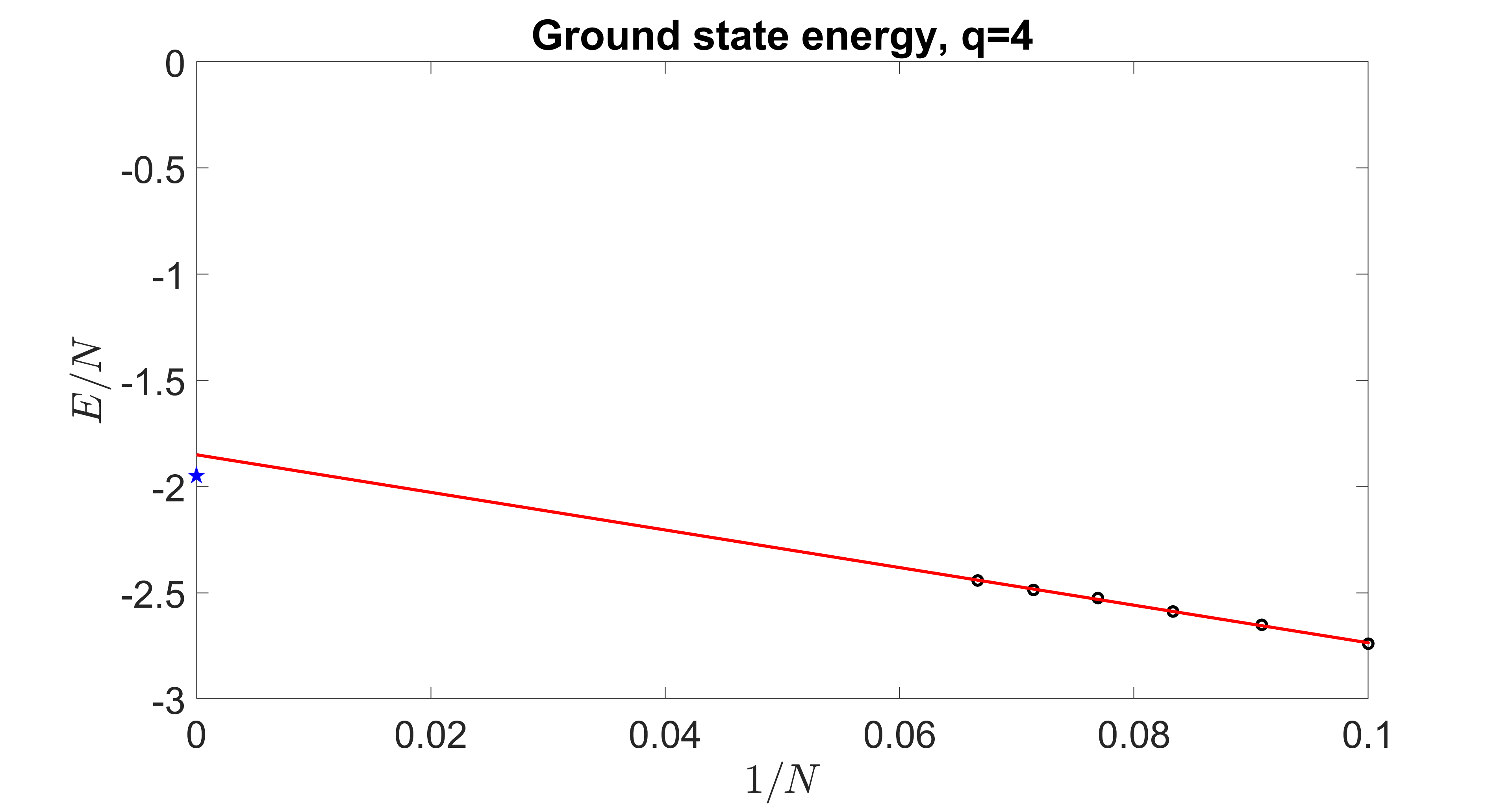}
    \caption{The energy per spin plotted versus $1/N$ for $q=4$. The red line is a linear extrapolation to $1/N=0$ and the blue star is the large $q$ prediction. Each point is $500$ samples. }
    \label{fig:q4Egs}
\end{figure}

Next, let us consider $q=4$. This case is more complicated because the model has an anti-unitary time-reversal symmetry which takes $\sigma_{r\mu} \to -\sigma_{r\mu}$. For odd $N$, the states organize into degenerate doublets related by time-reversal, while for even $N$ there is no exact degeneracy. We generalize the $\q$ order parameter for even $q$ to take into account the lowest two states,
\begin{equation}
    q_{EA} = \frac{1}{2N} \sum_{r \mu} \sum_{ab=1,2} |\langle a | \sigma_{r\mu} |b \rangle|^2.
\end{equation}
The modified EA order parameter is shown in Figure~\ref{fig:q4Qea} where we see a slow decrease as a function of system size. This behavior is suggestive of a non-glassy ground state, although this case seems marginal compared to the robust decrease seen below for $q=5$. A non-glassy ground state is supported by the gap histogram in Figure~\ref{fig:level4} where we see level repulsion. Nevertheless, the asymptotic limit at large $N$ is less definite than for $q=3$ and $q=5$. We also show an extrapolation of the ground state energy in Figure~\ref{fig:q4Egs}. The proximity to the large $q$ result is synergistic with \eqref{eq:Ederiv} and our observation that the large $q$ ansatz works well near $\tau=0$.

\begin{figure}
    \centering
    \includegraphics[width=.8\textwidth]{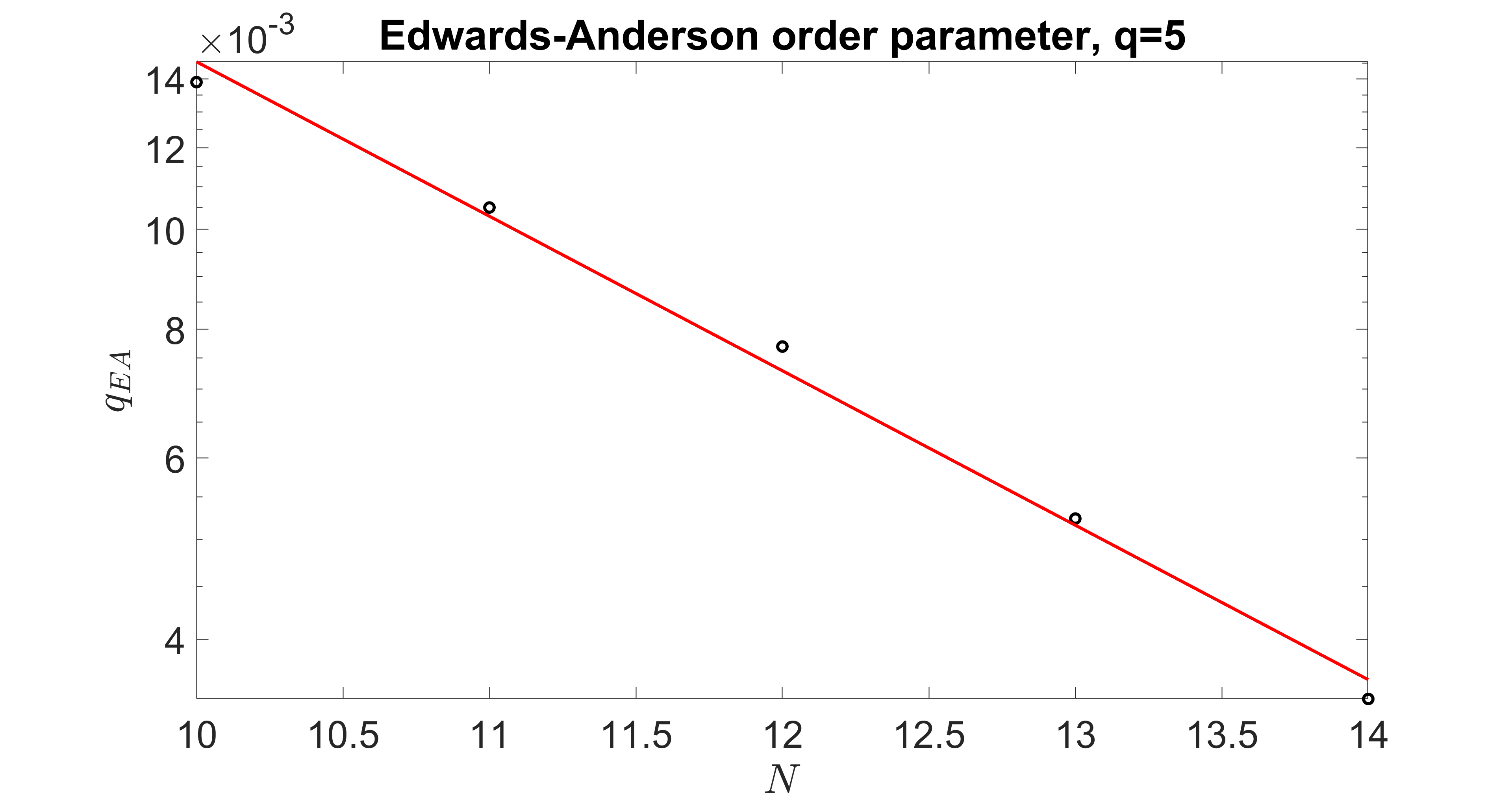}
    \caption{The Edwards-Anderson order parameter in the ground state (log scale) for $q=5$ as a function of system size. Each point is averaged over $2000$ samples. The rapid decay with system size strongly indicates the absence of glassiness. The red line is a linear fit to $\log \q$.}
    \label{fig:q5Qea}
\end{figure}

\begin{figure}
    \centering
    \includegraphics[width=.8\textwidth]{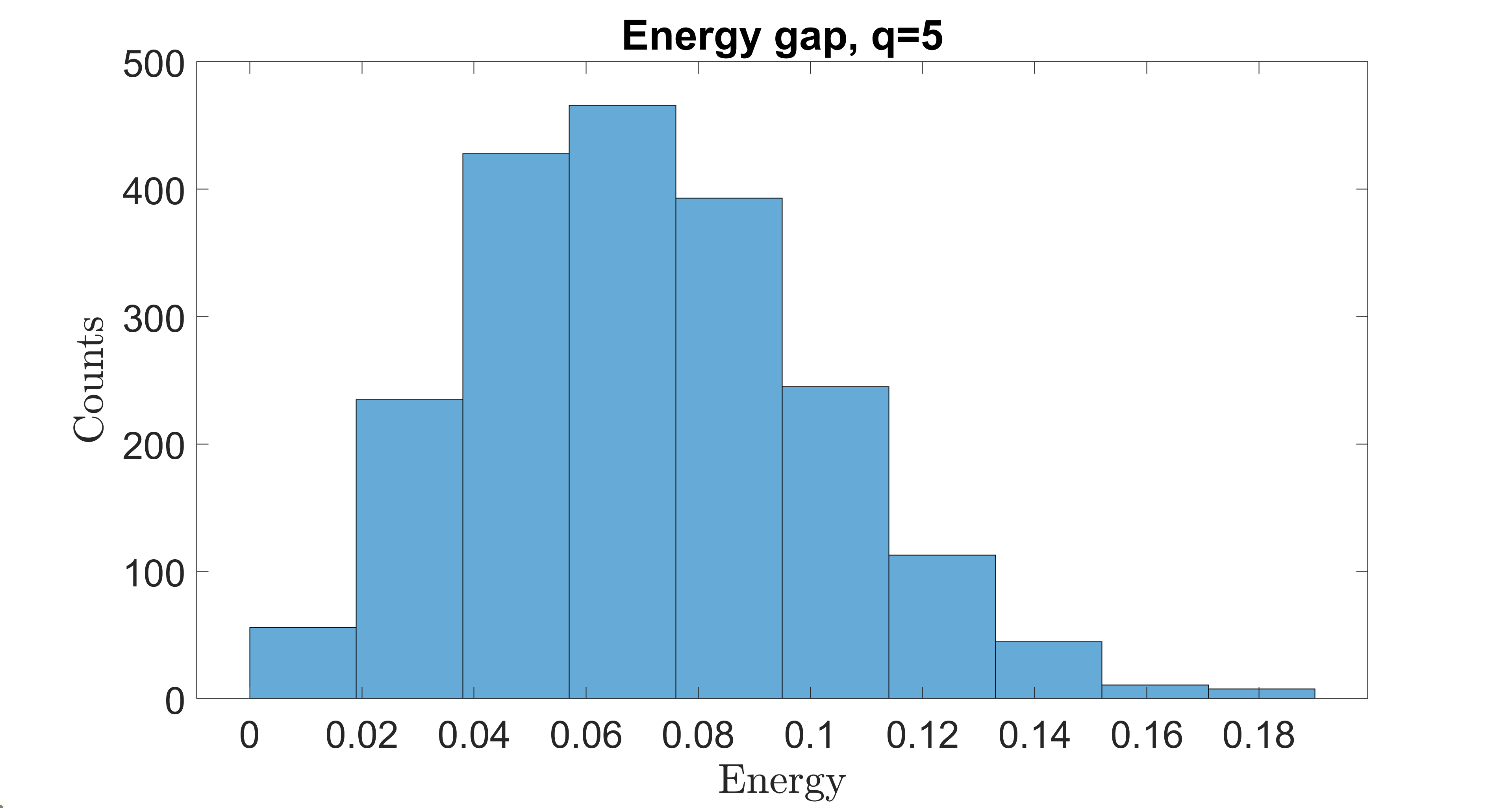}
    \caption{A histogram of $2000$ samples of the finite size energy gap between the ground state and first excited state for $q=5$ and $n=14$. The clear level repulsion is consistent with a non-glassy low energy state.}
    \label{fig:level5}
\end{figure}

\begin{figure}
    \centering
    \includegraphics[width=.8\textwidth]{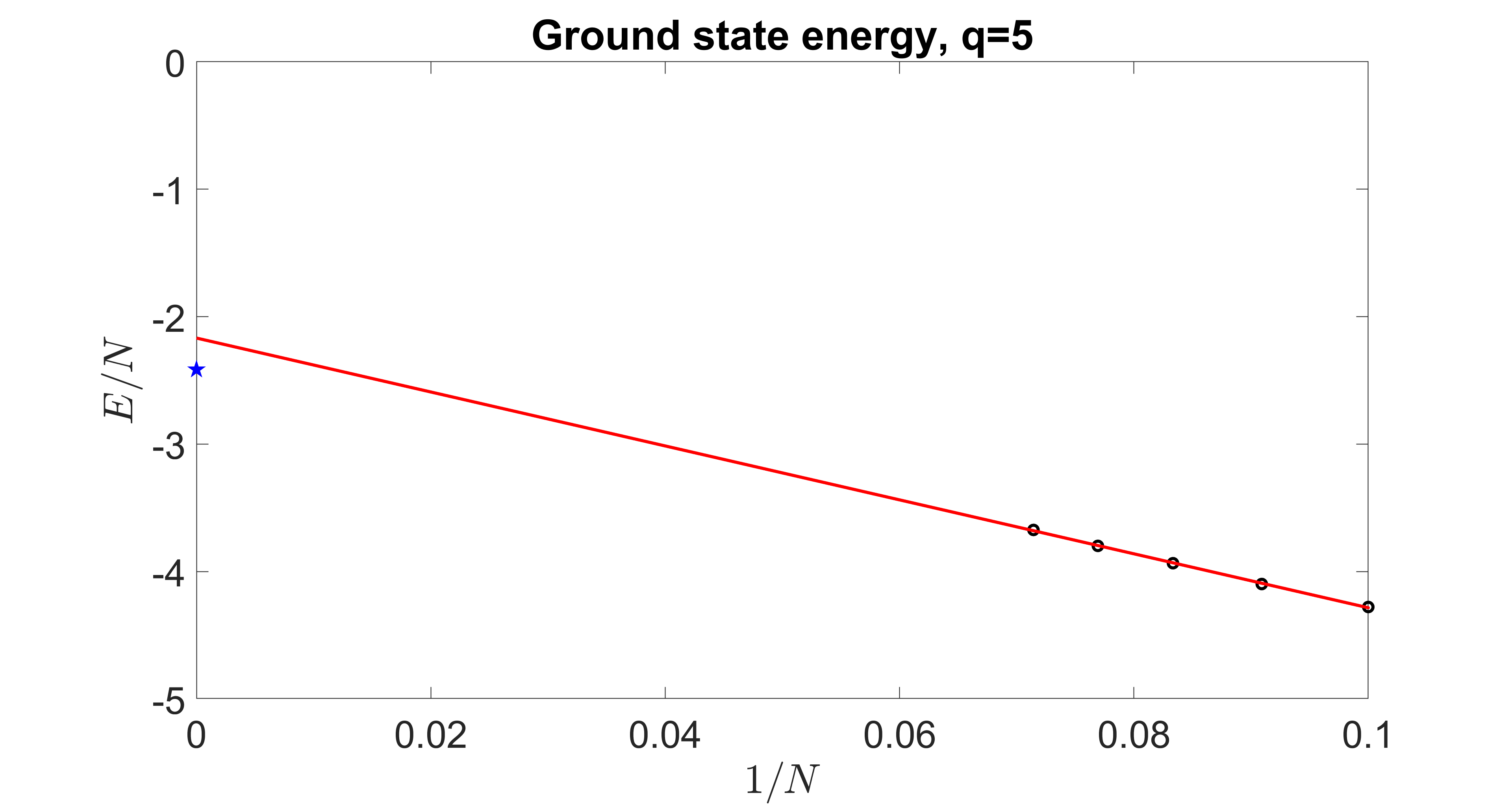}
    \caption{The energy per spin plotted versus $1/N$ for $q=5$. The red line is a linear extrapolation to $1/N=0$ and the blue star is the large $q$ prediction. Each point is $2000$ samples. }
    \label{fig:q5Egs}
\end{figure}

Finally, let us consider $q=5$. The EA order parameter is shown in Figure~\ref{fig:q5Qea} where we see an exponential decrease as a function of system size. This indicates that the $q=5$ ground state is not a glass. This conclusion is supported by the gap histogram in Figure~\ref{fig:level5}. This is exactly what we would expect if the ground state and the first excited state had random-matrix-like level repulsion. We also show an extrapolation of the ground state energy in Figure~\ref{fig:q5Egs}. The proximity to the large $q$ result is synergistic with \eqref{eq:Ederiv} and our observation that the large $q$ ansatz works well near $\tau=0$.

\begin{figure}
    \centering
    \includegraphics[width=.8\textwidth]{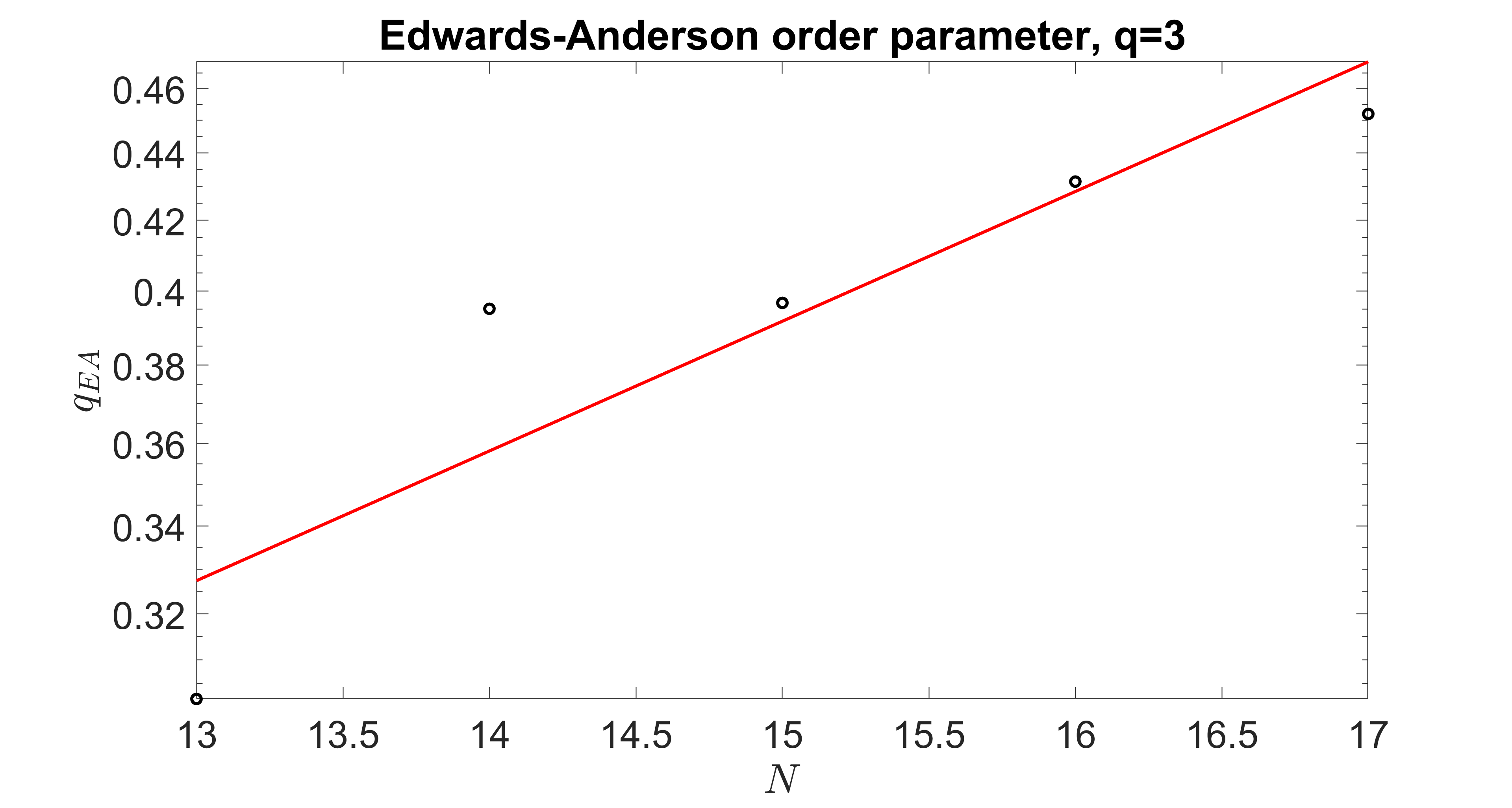}
    \caption{The Edwards-Anderson order parameter in the ground state (log scale) for $q=3$ as a function of system size from $N=13$ to $N=17$. Each point is averaged over $100$ samples. The approximate saturation with system size indicates a glassy ground state at large $N$. The red line is a linear fit to $\log \q$.}
    \label{fig:q3Qea_N17}
\end{figure}

\begin{figure}
    \centering
    \includegraphics[width=.8\textwidth]{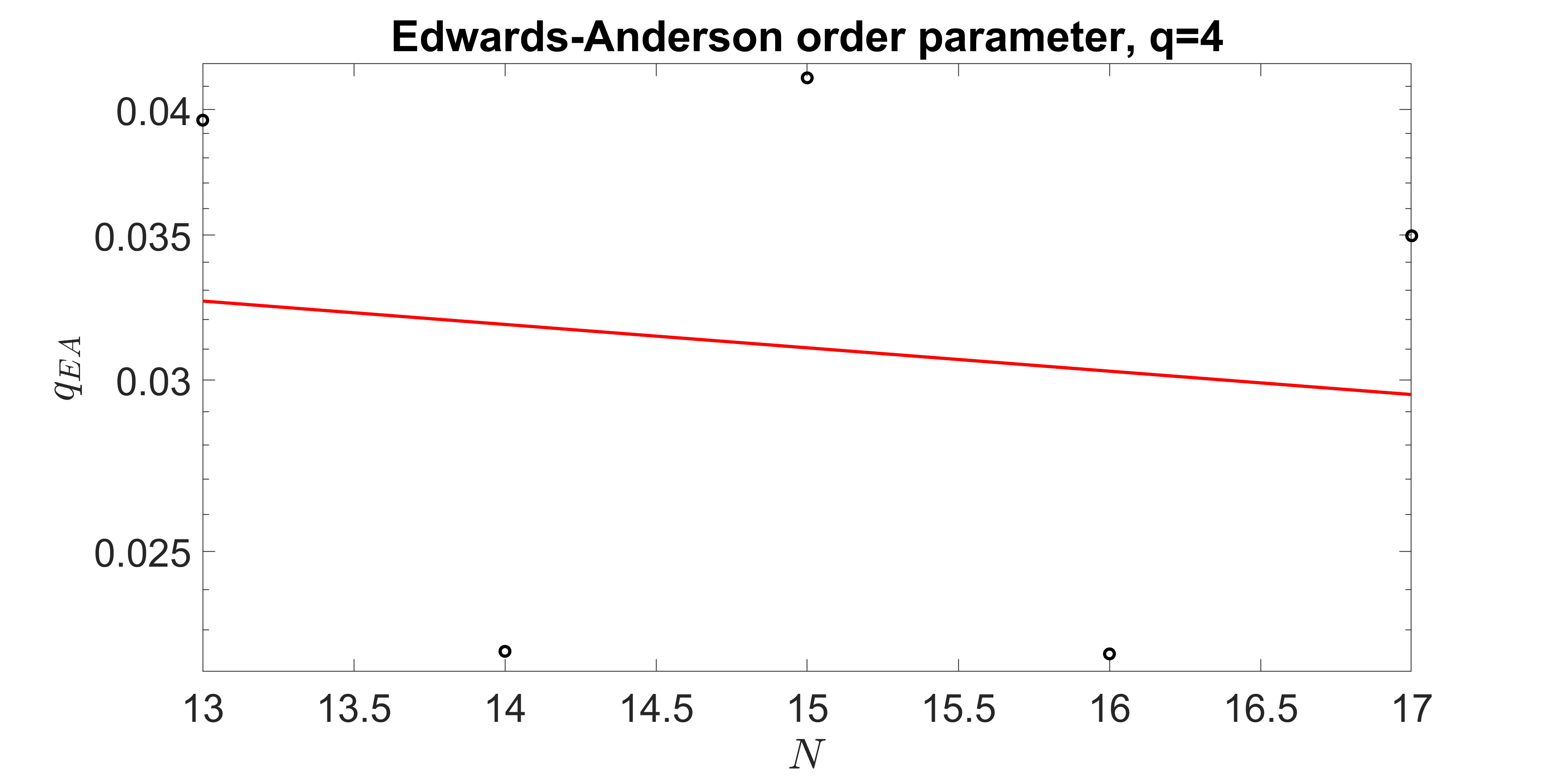}
    \caption{The Edwards-Anderson order parameter in the ground state (log scale) for $q=4$ as a function of system size from $N=13$ to $N=17$. Each point is averaged over $100$ samples. The approximate saturation with system size indicates a glassy ground state at large $N$. The red line is a linear fit to $\log \q$.}
    \label{fig:q4Qea_N17}
\end{figure}

\begin{figure}
    \centering
    \includegraphics[width=.8\textwidth]{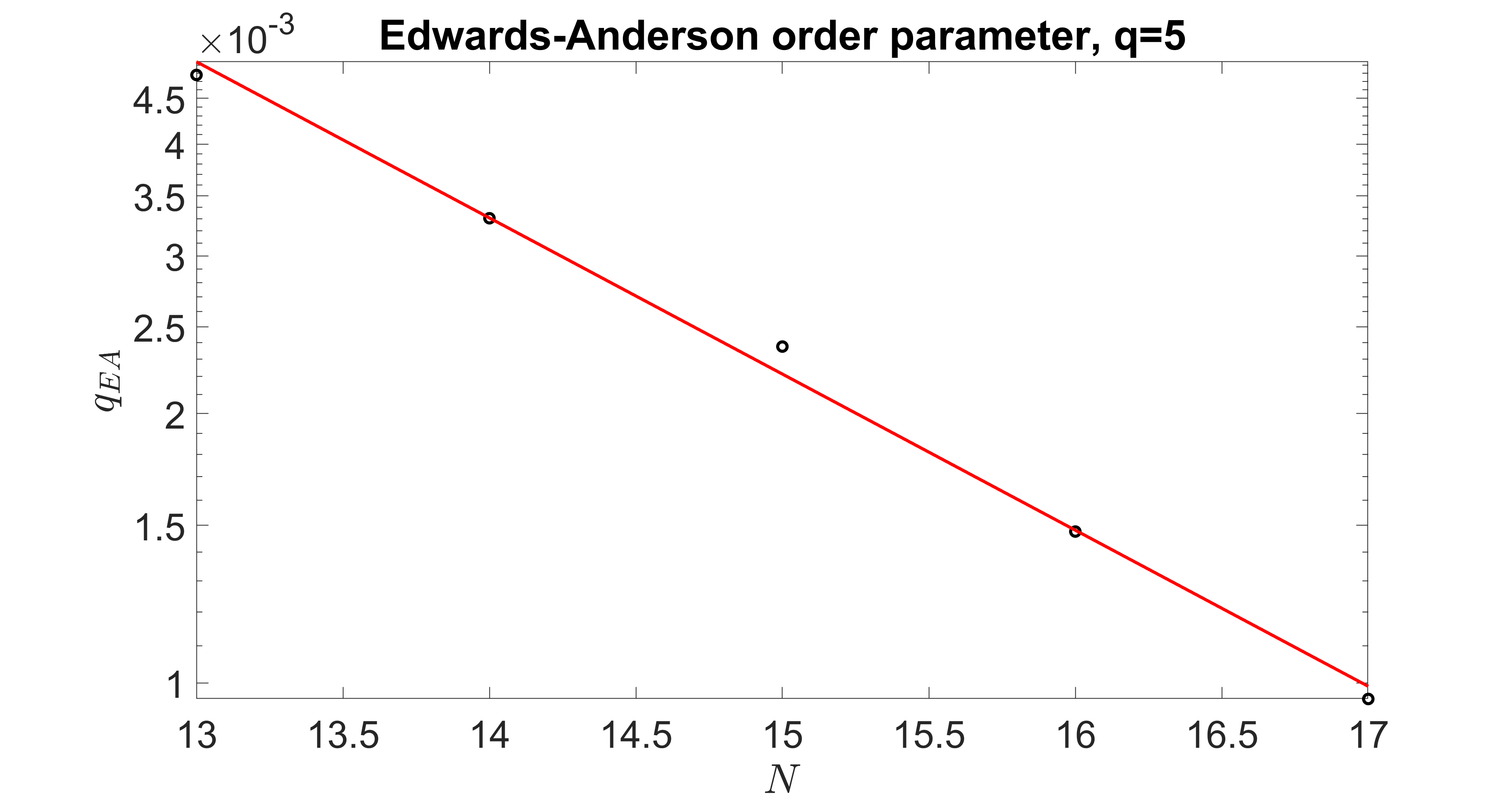}
    \caption{The Edwards-Anderson order parameter in the ground state (log scale) for $q=5$ as a function of system size from $N=13$ to $N=17$. Each point is averaged over $100$ samples. The rapid decrease with system size indicates a non-glassy ground state at large $N$. The red line is a linear fit to $\log \q$.}
    \label{fig:q5Qea_N17}
\end{figure}

\begin{figure}
    \centering
    \includegraphics[width=.8\textwidth]{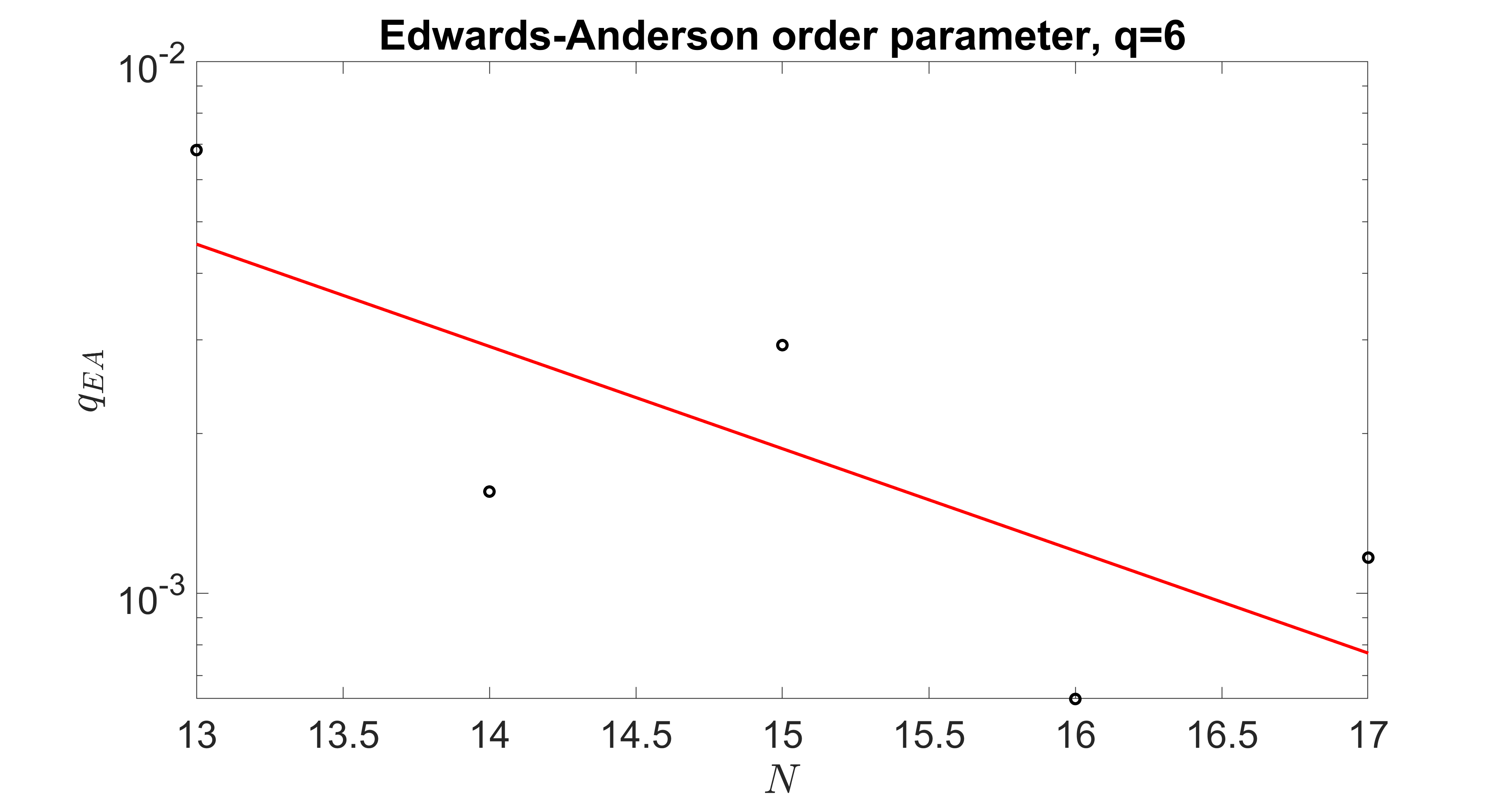}
    \caption{The Edwards-Anderson order parameter in the ground state (log scale) for $q=6$ as a function of system size from $N=13$ to $N=17$. Each point is averaged over $100$ samples. The rapid decrease with system size indicates a non-glassy ground state at large $N$. The red line is a linear fit to $\log \q$.}
    \label{fig:q6Qea_N17}
\end{figure}

\begin{figure}
    \centering
    \includegraphics[width=.8\textwidth]{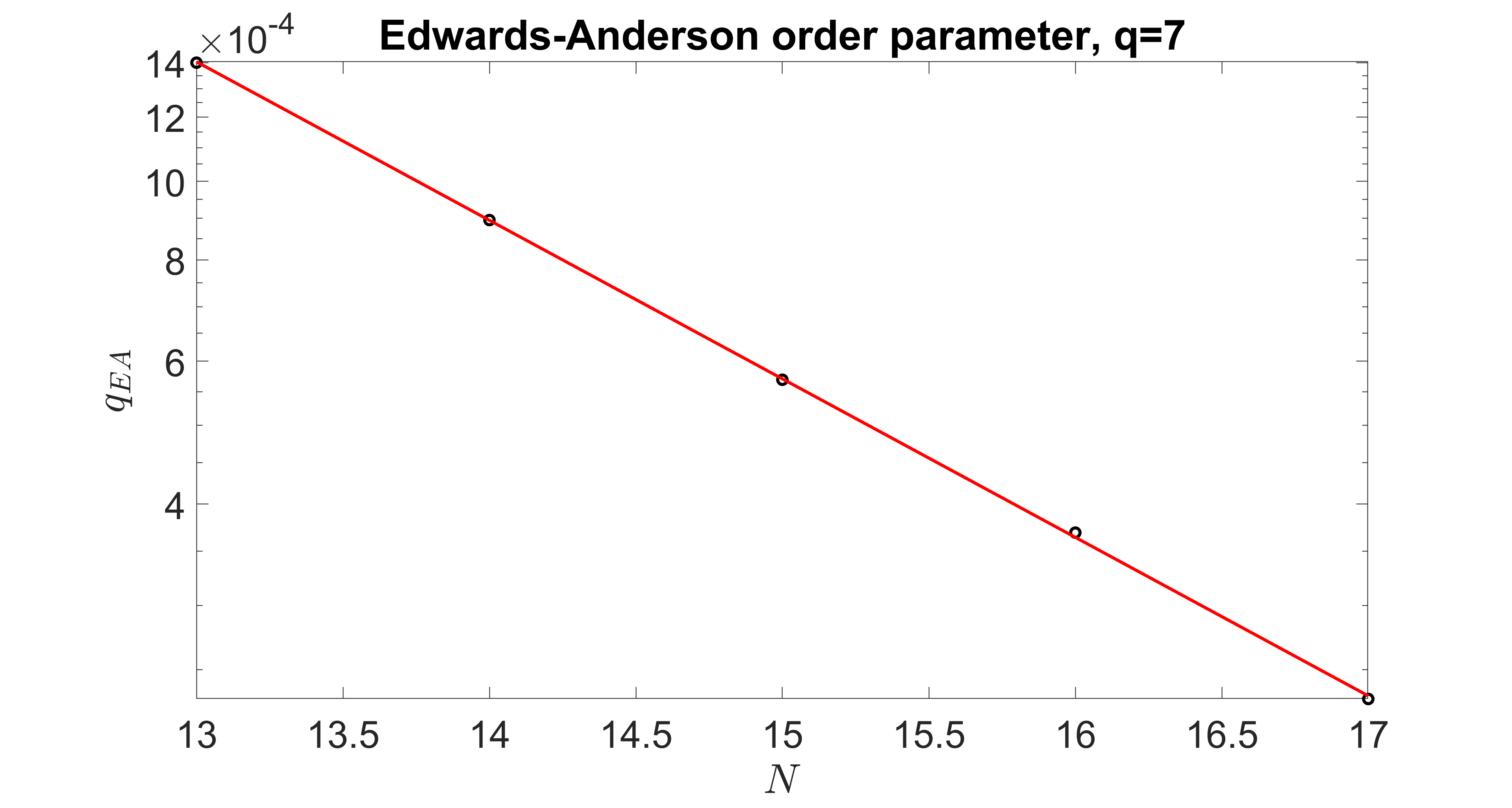}
    \caption{The Edwards-Anderson order parameter in the ground state (log scale) for $q=7$ as a function of system size from $N=13$ to $N=17$. Each point is averaged over $100$ samples. The rapid decrease with system size indicates a non-glassy ground state at large $N$. The red line is a linear fit to $\log \q$.}
    \label{fig:q7Qea_N17}
\end{figure}

To shed further light on the physics, and in particular to try to understand the fate of $q=4$ at large $N$, we carried out a second computation at somewhat larger system size, up to $N=17$, with $100$ samples per point. For these data we set the sparsity parameter to $\kappa=16$. The results for the EA order parameter for $q=3,4,5,6,7$ are shown in Figures \ref{fig:q3Qea_N17}, \ref{fig:q4Qea_N17}, \ref{fig:q5Qea_N17}, \ref{fig:q6Qea_N17}, and \ref{fig:q7Qea_N17}. At these somewhat larger sizes, we see indications of saturation for $q=3,4$ and robust exponential decays with system size for $q\geq 5$.

Our conclusion from this section is that $q=3$ is glassy, $q=4$ is likely glassy, and $q\geq 5$ shows no sign of glassiness. Although we did not study $q>5$ in detail, based on all the accumulated evidence and trends, we conjecture that $q\geq 5$ are all non-glassy and SYK-like down to the lowest temperatures. For $q=3,4$, the relatively small values of $\q$ suggest the existence of a wide intermediate energy window where the physics could still be SYK-like even if the system is glassy at asymptotically low temperature.

\pagebreak

\section{The \gsyk{} Model}
\label{sec:q_col}

To gain additional analytical insight into the physics, this section studies a generalization of \name{} that we call the \gsyk{} model. This model is indexed by an even integer $M=4,6,\cdots$ such that $M=4$ is \name{} and the model is solvable at large $M$. The model is very analogous to the SU($M$) magnets studied by Sachdev and Ye~\cite{SY}, suitably modified since spin-rotation symmetry plays no role for us. We emphasize that the model is bosonic for any $M$, but it is convenient to construct it using fermions. We consider $N$ sites each of which contains $M$ Majorana fermions. These fermions are denoted $\chi_{r\alpha}$, with $r=1,\cdots,N$ and $\alpha =1,\cdots,M$, and they obey
\begin{equation}
    \{\chi_{r\alpha},\chi_{r'\alpha'}\} = \delta_{rr'}\delta_{\alpha \alpha'}.
\end{equation}
For each site, we have a fermion parity operator,
\begin{equation}
    (-1)^{F_r} = (-2i)^{M/2} \chi_{r1} \cdots \chi_{rM}. 
\end{equation}
The physical Hilbert space is the even parity sector,
\begin{equation}
    (-1)^{F_r}=1,
\end{equation}
for each site $r$. The local Hilbert space dimension is thus $d = 2^{M/2-1}$.

From each site we can construct fermion bilinears,
\begin{equation}
    O_{r,\alpha\beta}=- 2i \chi_{r\alpha} \chi_{r\beta},
\end{equation}
which are bosonic operators that commute with all the $(-1)^{F_r}$s and square to the identity. The Hamiltonan is built from these operators as
\begin{equation}
    H = \sum_{r_1 \alpha_1 \beta_1\cdots r_q \alpha_q \beta_q} J_{r_1 \alpha_1 \beta_1\cdots r_q \alpha_q \beta_q} O_{r_1, \alpha_1 \beta_1} \cdots O_{r_q,\alpha_q \beta_q}.
\end{equation}
The $J$ couplings have variance
\begin{equation}
    \av{J_{\cdots}^2} = \frac{(q-1)! J^2}{N^{q-1}} \left[ \frac{\binom{M}{2}^q 2^{2q+1}}{M} \right]^{-1}.
\end{equation}
The bizarre numerical factor in square brackets is chosen to make the equations of motion simple.

In the case $M=4$, one may check that the $O_{r,\alpha \beta}$, of which there are six per site, can be identified with the Pauli matrices within the two-dimensional even parity subspace. Hence, apart from the fact that each Pauli appears twice in the sum over $\alpha<\beta$ (e.g. $O_{12}=O_{34}$ on the physical Hilbert space), we obtain exactly the Hamiltonian of \name{} when $M=4$. This doubling of the two-fermions operators is a consequence of the fact that $M=4$, and the four-fermi operator $(-1)^{F_r}$ is one on the physical Hilbert space.

We now study the model at large $M$ and show that it is solvable using standard SYK technology. In fact, this model is a generalization of coupled cluster models (e.g.~\cite{bentsen_scramblers_sparse}) in which we allow for $q$-body interactions between the clusters/sites and we gauge the fermion parity on each cluster/site. The result is a purely bosonic model in the sense discussed above: the total Hilbert space is a tensor product of local Hilbert spaces and operators on different tensor factors commute. The fermions $\chi_{\alpha}$ are analogous to ``partons'' or ``spinons'' in the condensed matter language, and they offer a convenient way to describe the operators of interest, the $O_{\alpha \beta}$.

We discuss the single replica path integral for \gsyk{} in two stages, first for general $M$ and then at large $M$. Using fermion coherent states, the path integral for general $M$ is
\begin{equation}
    \av{Z} = \int \CD^{NM} \chi \left\{ \prod_r \frac{1+(-1)^{F_r}}{2}\right\}\exp\left( - \sum_{r\alpha} \frac{1}{2} \int d\tau \chi_{r\alpha} \partial_\tau \chi_{r\alpha} + \frac{J^2 }{2 q N^{q-1}} \frac{M}{\binom{M}{2}^q 2^{2q+1}} \sum_{r_1 ,\alpha_1 < \beta_1 \cdots} \left[ \int d\tau O_{r_1, \alpha_1 \beta_1} \cdots \right]^2 \right).
\end{equation}
The factor in curly braces is a projector which restricts to even fermion parity on each site. We then decouple the interaction term by introducing a Lagrange multiplier term,
\begin{equation}
    \frac{M N}{16} \int d\tau d\tau' \Sigma(\tau,\tau') \left[ G(\tau,\tau') - \frac{1}{N \binom{M}{2}}  \sum_{r, \alpha<\beta} O_{r,\alpha \beta}(\tau) O_{r,\alpha \beta}(\tau') \right],
\end{equation}
which allows the path integral to be written as 
\begin{equation}
    \av{Z} = \int \CD G \CD \Sigma e^{- N I}
\end{equation}
with
\begin{equation}
    I = \int d\tau d\tau' \left[ - \frac{J^2 }{2 q }\frac{M}{2^{2q+1}}  G(\tau,\tau')^{q} + \frac{M}{16} \Sigma(\tau,\tau') G(\tau,\tau') \right] + I_{\text{site}}
\end{equation}
and
\begin{equation}
    I_{\text{site}} = - \log \left[ \int \CD^M \chi \left\{ \frac{1+(-1)^F}{2}\right\} \exp\left( - \sum_\alpha \frac{1}{2} \int d\tau \chi_\alpha \partial_\tau \chi_\alpha + \frac{M}{16 \binom{M}{2}} \sum_{\alpha<\beta} \int d\tau d\tau' \Sigma(\tau,\tau') O_{\alpha \beta}(\tau)O_{\alpha \beta}(\tau') \right) \right].
\end{equation}

Before proceeding to the large $M$ limit, let's understand the effect of the fermion parity projector. The effect of the $(-1)^F$ insertion is to change the boundary conditions of the fermions around the imaginary time circle from anti-periodic (AP) to periodic (P). Hence, the full single site path integral is a sum of two terms, 
\begin{equation}
    e^{-I_{\text{site}}} = \frac{Z_{\text{AP}} + Z_{\text{P}}}{2},
\end{equation}
where $Z_{\text{AP}}$ is a conventional term  without $(-1)^F$ and $Z_{\text{AP}}$ is a twisted term $Z_{1,P}$ with $(-1)^F$. In the large $M$ limit, this is not expected to change the physics of the model significantly, meaning that the free energy per fermion should be the same in the even and odd parity sectors, up to $1/M$ corrections. We first describe the large $M$ limit for the untwisted contribution, which counts both fermion parity sectors. Then we discuss how the twisted contribution is obtained.

The point of the large $M$ limit is that it allows us to evaluate the path integral over $\chi_\alpha$ in the definition of $I_{\text{site}}$ by saddle point. We introduce another Lagrange multiplier term,
\begin{equation}
    \frac{ M }{2} \int d\t d\t' \Sigma_\chi(\t,\t') \left[ G_\chi(\t,\t') - \frac{1}{M} \sum_\alpha  \chi_{\alpha}(\tau) \chi_\alpha(\tau') \right],
\end{equation}
so the single site path integral becomes
\begin{equation}
    Z_{\text{AP}} = \int \CD G_\chi \CD \Sigma_\chi \exp\left( M \log \text{Pf}(\partial_\tau - \Sigma_\chi) + \frac{M}{2} \int d\t d\t' \left[ \frac{1}{2} \Sigma(\t,\t') G_\chi(\t,\t')^2 - \Sigma_\chi(\t,\t') G_\chi(\t,\t') \right] \right).
\end{equation}

The EOM for $\Sigma_\chi$ and $G_\chi$ are, respectively, 
\begin{equation}
    G_\chi = \frac{1}{\partial_\tau - \Sigma_\chi}
\end{equation}
and
\begin{equation}
    \Sigma_\chi = \Sigma G_\chi.
\end{equation}
The EOM for $G$ and $\Sigma$ are, respectively,
\begin{equation}
    \Sigma = \frac{J^2 }{2^{2q-2}} G^{q-1} = J^2 \left(\frac{G}{4} \right)^{q-1}
\end{equation}
and
\begin{equation}
    G = 4 G_\chi^2.
\end{equation}
These equations can be reduced to a closed set involving just $G_\chi$ and $\Sigma_\chi$, 
\begin{equation}
    G_\chi = (\partial_\tau - \Sigma_\chi)^{-1},
\end{equation}
\begin{equation}
    \Sigma_\chi = J^2  G_\chi^{2q-1}.
\end{equation}
From this we learn that $G_\chi$ is the correlator of an SYK model with $q_{\chi}=2q$.  

What about the twisted contribution, $Z_{\text{P}}$? Starting with a solution $G(\t,\t')$ of the untwisted path integral, we can obtain a solution of the twisted EOM by modifying $G$ such that it picks up an extra minus sign whenever $\t$ or $\t'$ move past $\beta$. The reader may worry that this choice of time breaks time-translation symmetry, but it is actually a gauge artifact and time-translation symmetry is preserved for any gauge invariant observable. This twisted $G$ is $\tilde{G}(\t,\t') = Q(\t) G(\t,\t') Q(\t')$ for some $Q$ that implements the twist, e.g. $Q=\text{sgn}(\beta-\tau)$ for $\tau$ near $\beta$. Since this is effectively a pre- and post-multiplication of $G$ by a diagonal matrix, $Q$, that squares to the identity, it follows that
\begin{equation}
    \text{Pf}(\tilde{G}^{-1}) = \text{Pf}( Q G^{-1} Q) = \text{Pf}( G^{-1})
\end{equation}
and
\begin{equation}
    \int \tilde{G}^q = \int G^q, \int \tilde{G} \tilde{\Sigma} = \int G \Sigma.
\end{equation}
Hence, the value of the saddle point action will be identical. Now, it may still happen that the full twisted contribution vanishes, but we at least learn that $Z_{\text{P}} = c Z_{\text{AP}}$ for some value $c$ (possibly zero) which is only polynomially large in $M$ (and hence cannot change the leading order free energy per fermion at large $M$).

What all this means is simply that, at large $M$, we can ignore the fermion parity projector. We can therefore simply stick with the standard anti-periodic contribution to $e^{-I_{\text{site}}}$. Considering the EOM and going to the conformal regime, we learn that the Majorana field has dimension $\Delta_\chi= 1/q_{\chi} = 1/(2q)$. The ``spin'' correlator $G$ is the square of the fermion correlator, so in the conformal regime, it will be proportional to the conformal result with dimension $\Delta=1/q$. This is precisely what we found numerically for $M=4$ and $q\geq 3$. Note that in the case $q=2$, we do get SYK-like physics over a wide range of energies at large $M$, but the SYK-like saddle suffers from an known instability which results in glassy physics at the lowest energies~\cite{SY,haehl_crossover}.

\subsection{Bounds on Glassiness at Large $M$}

What happens to the system at large but finite $M$? From our numerics on \name{} in Sec.~\ref{sec:ed}, it seems likely that the model is just SYK-like all the way down to $M=4$ for sufficiently large $q$. Still, we would like to constrain possible glassy states.

We can derive a bound on the analog of the Edwards-Anderson order parameter for \gsyk{} (similar to the recent analysis in \cite{haehl_crossover}). With the normalization of $G$ chosen above, we have $G(0)=1$. Now introducing replicas $a=1,\cdots,n$, we have $G_{11}(0)=1$ and we ask how large can $G_{1b}$ be? This off-diagonal term can be bounded by constraining the possible expectation values of the $O_{\alpha\beta}$. The best we can do is choose a state of the fermions in which 
\begin{equation}
    - 2 i \chi_1 \chi_2 = - 2 i \chi_3 \chi_4 = \cdots = -2 i \chi_{M-1} \chi_M = 1,
\end{equation}
or some permutation thereof. In such a state, $M/2$ of the $\binom{M}{2}$ fermion bilinears have a non-zero expectation value, so 
\begin{equation}
    G_{1 b} \leq \frac{1}{M-1}.
\end{equation}
This formula tells us that glassy physics is suppressed at large $M$. For $M=4$, this formula reduces to the previous result that $G_{1b}/G_{11}(0) \leq 1/3$. 

Suppose that we are at sufficiently large $M$ or $q$ such that $G$ can develop into its conformal form for some range of $\beta$. $G$ is smallest at $\tau=\beta/2$ where it is
\begin{equation}
    G(\beta/2) \sim 1/(\beta J)^{2/q},
\end{equation}
which only approaches $1/M$ when 
\begin{equation}
    \beta J \sim M^{q/2}.
\end{equation}
At large $M$ and/or large $q$, this indicates the existence of a parametrically large window in energy over which the physics can be SYK-like. 

\subsection{Comments on Generic $q$-Local Interactions }

We can further generalize the \gsyk{} model by expanding the set of local operators which can be appear in the $q$-body interaction. So far, we restricted these operators to be fermion bilinears, but we could allow more general operators, such as $\chi_\alpha \chi_\beta \chi_\gamma \chi_\delta$. In fact, if we allow all possible even powers of fermions, we obtain a complete basis for operators on the $2^{M/2-1}$-dimensional site. We make two comments about this more generic situation.

First, while it is more complicated to address the case where we add all possible operators with random couplings of equal variance, it is easier to discuss adding additional operators perturbatively, i.e. with a random coupling whose variance is suppressed by a small factor. From the fermion perspective, this looks like adding higher-$q_\chi$ interactions. But such higher-$q_\chi$ interactions are typically irrelevant in the renormalization group sense. This means we expect the lowest $q_\chi$ present in the Hamiltonian to control the physics at very low energy. Hence, it could be that adding these terms actually does not strongly modify the very low temperature physics. In other words, adding generic interactions perturbatively could preserve the SYK-like physics when it is present at very low energy.

Second, as we add more operators to the Hamiltonian, the corresponding bound on glassiness improves. For example, suppose we allow every possible single-site operator in the $q$-body interactions. There are $d^2-1$ such operators (the number generators of SU($d$) and the number of distinct operators built from even products of fermions), where again $d = 2^{M/2-1}$. Let us also continue to normalize $G$, which now includes a sum over all single site operators, so that $G(0)=1$. Using the same fermion state as above, we learn that $d-1$ distinct operators now have a non-vanishing expectation value. This number is obtained by considering general operators of the form $(O_{12})^{x_1} (O_{34})^{x_2} \cdots$, accounting for a double counting thanks to $(-1)^F=1$, and subtracting one for the identity. Hence, we have
\begin{equation}
    G_{1b} \leq \frac{d-1}{d^2-1} = \frac{1}{d+1},
\end{equation}
which again reduces to $G_{1b}/G_{11}(0) \leq 1/3$ for $d=2$ ($M=4$).

\section{Physics of the SYK-like Saddle}
\label{sec:syk-like}

Here we briefly discuss the low-energy physics when the power-law decaying $G(\tau)$ controls the quenched free energy. In particular, in SYK the power-law decay of the Green function is just one part of a whole suite of interrelated phenomena. Do we obtain the same set of phenomena in \name{}?

The general idea is to appeal to the existence of approximate time-reparameterization symmetry. The first two terms in \eqref{eq:GSigma} are manifestly invariant under reparameterizations, but the third term breaks this symmetry explicitly. The ground state solution also breaks the symmetry spontaneously, thus yielding a pattern of symmetry breaking familiar from SYK. 

For example, under a general reparameterization $\t \to f(\t)$, $\Sigma$ is modified to
\begin{equation}
    \Sigma(\tau_1,\tau_2)\to f'(\tau_1)^{1-\Delta} f'(\tau_2)^{1-\Delta} \Sigma(f(\tau_1),f(\tau_2))
\end{equation}
with $\Delta=1/q$. Every choice of $h$ now has a reparameterized form, 
\begin{equation}
    h(\tau) \rightarrow f'(\tau)^{1-\Delta} h(f(\t)).
\end{equation}
If we could rescale the integration variables $s(\t)$ to $f'(\tau)^\Delta s(f(\tau))$, then the $G$ function would transform as expected. Of course, this is strictly speaking impossible since $s(\tau)$ lives on the sphere and there are additional microscopic terms in the measure. However, after an appropriate coarse-graining, the coarse-grained $s(\tau)$ is no longer confined to the sphere. This suggests that the failure of time reparameterization symmetry is indeed a UV issue.

This means that when we evaluate that action on a reparameterized configuration,
\begin{equation}
    f'(\t_1)^\Delta f'(\t_2)^\Delta G(f(\t_1),f(\t_2)),
\end{equation}
the contribution comes from the UV terms which explicitly break the symmetry. Following the usual analysis, we expect the Schwarzian term to be among these contributions. If this is the leading term in the action, then the physics is very close to that of SYK.

Another possibility at finite $q$ and $M$ is that there could be weakly irrelevant operators which contribute a non-local action that dominates over the Schwarzian~\cite{milekhin2021coupled}. This is a possibility when the theory has an irrelevant operator $O$ with dimension $\Delta_O \in (1,3/2)$ that appears in the microscopic action~\cite{maldacena_nearly_ads2}. Whether this possibility is realized depends on the operator spectrum and on the nature of the UV terms. Even for models where this possibility is realized, we still have an extensive ground state entropy, maximal chaos, and the Schwarzian is still present~\cite{milekhin2021coupled}.

For example, take $q=3$. This case is glassy for \name{}, but for sufficiently large $M$ in \gsyk{}, the SYK-like solution should control the quenched free energy over a wide range of temperatures. In this case, $\Delta =1/3$ and the corresponding long-range fixed point is in the Gaussian regime. For a Gaussian fixed point, we know the dimensions of all the operators in the theory and there are seemingly operators with dimensions in the range $\Delta_O \in (1,3/2)$, e.g. $\phi^4$ where $\phi$ is some coarse-grained form of $O_{\alpha \beta}(\t)$. Nevertheless, it is not clear if these operators appear in the UV, especially since they do not contain time-derivatives whereas the coherent state overlaps in the measure are unity unless the fields vary in time.

These general comments can be elaborated in the various solvable limits we considered in the paper. Consider first the large $q$ limit. From the discussion in Section \ref{sec:Liouville} and the numerical solutions of the EOM in Section \ref{sec:Numerics}, we know the SYK-like saddle in this regime hews close to the literal SYK solution, up to a rescaling of $G$ and a different numerical prefactor relating $J$ and $\mathcal{J}$. Thus at large $q$ we expect the thermodynamics to essentially match those of the large $q$ SYK model. This means an SYK-like extensive zero temperature entropy which approaches the infinite temperature entropy as $q$ becomes large. We also should have the large $q$ chaos properties, including maximal chaos at low temperature. Moreover, in the scaling limit in which $q^2/N$ is held fixed as $N\to \infty$, SYK is solvable~\cite{berkooz_dssyk} and \name{} will have the same physics~\cite{berkooz_chord_sg,berkooz_beyond_epr}.

Similarly, at large $M$ in the \gsyk{} model, we know that the spin correlation $G$ is determined from $G_\chi$, which is exactly the correlator of a fermionic SYK with $q_\chi=2q$. From \eqref{eq:Eint}, suitably adjusted to account for the different normalization of $G$, we learn that
\begin{equation}
    - \frac{E}{N} \propto \int d\t (G_\chi(\t))^{2q},
\end{equation}
which in the conformal regime is identical to the SYK expression up to an overall factor. The integral is 
\begin{equation}
    - \frac{E}{N} - \text{(UV part)} \sim  \int_{\tau_c}^{\beta - \tau_c} d\t \frac{\pi^2}{\beta^2 \sin^2 \frac{\pi \tau}{\beta}} \sim - \frac{2 \pi}{\beta} \cot \frac{\pi \tau_c}{\beta}
\end{equation}
for some UV cutoff $\tau_c \sim 1/J$, which yields $E/N = E_0/N + c' T^2 + \cdots$.

\section{Discussion}
\label{sec:discussion}

In this work, we argued for the existence of SYK-like dynamics in a variety of bosonic models. Focusing on \name{}, we catalogued possible solutions to the single-replica EOM and then, in lieu of carrying out the minimax procedure, we used small scale numerical simulations to provide input on the nature of the quenched saddle. Additional analytical insight was provided by a large $q$ limit. We also introduced and analyzed the \gsyk{} model, showing that it has SYK-like behavior for sufficiently large $M,q$. Our numerical results suggest that this behavior persists down to $q=5$ in \name{}. Even when the asymptotically low-temperature physics is glassy, e.g. $q=2,3,4$, we also gave evidence for an intermediate regime of energies with SYK-like physics.

We considered a three component model, but there is also interest in a two-component version as recently discussed in \cite{Hanada:2023rkf}. One can develop an analogous formulation for the two-component model, with the main difference being that the effective spin chain has two components (XY) instead of three (XYZ). We expect the single-replica solutions that we found here will have an analog for the two-component action. Furthermore, we expect that the two-component model will also be SYK-like for sufficiently large $q$. We have preliminary results for the two-component case which we hope to present elsewhere.

These results have significant implications for the structure of quantum field theories, suggesting that nontrivial CFTs and holographic duals are more common than we thought. It also has clear implications for near-term quantum simulators, creating a more viable path to simulating SYK-like physics, and opening the door to analogous bosonic simulations of more complicated theories of gravity.

While this paper lays out the basics of $q$-local quantum physics in \name{}, there is much still to do. Some broad questions include:
\begin{enumerate}
    \item Are these results limited to 0+1D systems, or are there other situations where generic disordered systems have holographic duals? There has been work on systems with generic local interactions on a graph~\cite{ramis_generic} and in translation invariant spin chains~\cite{PhysRevB.100.035113}, but much remains to be understood.
    \item What can be said about fluctuations of the action away from the saddle? \name{} and its cousins have a reparameterization symmetry spontaneously broken down to $SL(2)$. Does that mean that they have a Schwarzian action that dominates? Not necessarily, as some systems \cite{milekhin2021coupled} have non-local effective actions. It should be possible to understand whether this is realized in \name{} by a careful study of the low-temperature thermodynamics and response functions.
    \item For the experimentalists / quantum simulationists, what is the actual two-point function for \name{} at $q=4$ at large system size? In section \ref{sec:Numerics} we find a replica-diagonal solution consistent with SYK-like physics exists but the exact diagonalization results suggest a small but non-zero value of $\q$. What about for \gsyk{} at $q=3$ for some modest intermediate $M$, such as $M=6$ or $8$? These questions are all ideal targets for a quantum simulator.
    \item For the holographers, since we know that quenched is not equal to annealed in \name{}, what is the gravitational analogue of the annealed solution (assuming the usual gravity solution is the analogue of the quenched solution)? How do these solutions interplay with the recent developments regarding the role of averaging in theories of gravity? Is there is holographic interpretation of the glassy state for $q=2,3,4$?
    \item Can we analyze the limit of large local Hilbert space dimension while also allowing generic local operators in the interaction? If so, is the physics SYK-like, at least over a wide range of energies, as suggested by the \gsyk{} model?
\end{enumerate}

Some more model-specific tasks include:
\begin{enumerate}
    \item It would be interesting to carry out a systematic study of the SYK-like solution at low energies to, for example, definitively establish the form of the heat capacity and to determine the zero temperature entropy. Perhaps the argument in~\cite{milekhin_entropy} can be useful.
    \item It would likewise be desirable to carry out a full numerical analysis of key observables in sparse finite $N$ instances, quantities such as the real-time decay of correlations and out-of-time-order correlations.
    \item More ambitiously, it would be very interesting to construct replica non-diagonal saddles and to carry out the minimax procedure. This is work in progress.
    \item It would also be very interesting to compute the leading $1/q$ correction~\cite{2018arXiv180106871T} in \name{} and the \gsyk{} model as well as the leading $1/M$ correction in \gsyk{}. 
\end{enumerate}

\appendix

\subsection*{Acknowledgements} We thank M. Hanada, A. Jevicki, X. Liu, E. Rinaldi, and M. Tezuka for many discussions. We also thank F. Haehl, C. Baldwin, S. Sachdev, and A. Milekhin for additional discussions and helpful feedback on the manuscript. We gratefully acknowledge support from Joint Quantum Institute and from AFOSR under grant number FA9550-19-1-0360. The numerical data were obtained using the Zaratan high-performance computing cluster at the University of Maryland, College Park. We also thank the Perimeter Institute and the It From Qubit Simons Collaboration for hosting a wonderful Last Hurrah where some of this work was done. 

\section{The \name{} Path Integral}
\label{app:PathIntegral}

\subsection{Spin Coherent States}

The path integral is derived using spin coherent states. Note that, although we use the conventional spin language, an instance of the model has no spin rotation symmetry. For a single spin-1/2 degree of freedom, we take the spin coherent state to be
\begin{equation}
    \ket{\Omega}= \begin{bmatrix}
        \cos \frac{\theta}{2} \\ \sin \frac{\theta}{2} e^{i \phi }
    \end{bmatrix}
\end{equation}
with
\begin{equation}
    \Omega = ( \sin \theta \cos \phi, \sin \theta \sin \phi, \cos \theta).
\end{equation}
The overlap of two spin coherent states is
\begin{equation}
    \langle \Omega | \Omega' \rangle = \cos \frac{\theta}{2} \cos \frac{\theta'}{2} +  \sin \frac{\theta}{2} \sin \frac{\theta'}{2} e^{- i (\phi - \phi')}.
\end{equation}

This set of states yields a resolution the identity ($d\Omega \equiv d\theta \sin \theta d\phi$):
\begin{equation}
   \int \frac{d\Omega}{2\pi} |\Omega \rangle \langle \Omega | = \int \frac{d\Omega}{2 \pi} \begin{bmatrix}
       \cos^2 \frac{\theta}{2} & \cos \frac{\theta}{2} \sin\frac{\theta}{2} e^{-i\phi} \\
       \cos \frac{\theta}{2} \sin\frac{\theta}{2} e^{i\phi} & \sin^2 \frac{\theta}{2}
   \end{bmatrix} = I.
\end{equation}
They can also be used to represent the Pauli matrices, 
\begin{equation}
   \int \frac{d\Omega}{2\pi} |\Omega \rangle \langle \Omega | (3 \Omega_x) = \int \frac{d\Omega}{2 \pi} \begin{bmatrix}
       \cos^2 \frac{\theta}{2} & \cos \frac{\theta}{2} \sin\frac{\theta}{2} e^{-i\phi} \\
       \cos \frac{\theta}{2} \sin\frac{\theta}{2} e^{i\phi} & \sin^2 \frac{\theta}{2}
   \end{bmatrix} (3 \sin \theta \cos \phi) = \sigma_x.
\end{equation}
\begin{equation}
   \int \frac{d\Omega}{2\pi} |\Omega \rangle \langle \Omega | (3 \Omega_y) = \int \frac{d\Omega}{2 \pi} \begin{bmatrix}
       \cos^2 \frac{\theta}{2} & \cos \frac{\theta}{2} \sin\frac{\theta}{2} e^{-i\phi} \\
       \cos \frac{\theta}{2} \sin\frac{\theta}{2} e^{i\phi} & \sin^2 \frac{\theta}{2}
   \end{bmatrix} (3 \sin \theta \sin \phi) = \sigma_y.
\end{equation}
\begin{equation}
   \int \frac{d\Omega}{2\pi} |\Omega \rangle \langle \Omega | (3 \Omega_z) = \int \frac{d\Omega}{2 \pi} \begin{bmatrix}
       \cos^2 \frac{\theta}{2} & \cos \frac{\theta}{2} \sin\frac{\theta}{2} e^{-i\phi} \\
       \cos \frac{\theta}{2} \sin\frac{\theta}{2} e^{i\phi} & \sin^2 \frac{\theta}{2}
   \end{bmatrix} (3 \cos \theta) = \sigma_z.
\end{equation}

To demonstrate the machinery, consider a single spin with Hamiltonian $H = -h \sigma_x$. For infinitestimal imaginary time $\Delta \tau$, we write
\begin{equation}
    e^{-\Delta \tau H} = I + \Delta \tau h \sigma_x = \int \frac{d\Omega}{2 \pi} |\Omega \rangle \langle \Omega | (1 + \Delta \tau h (3 \Omega_x)).
\end{equation}
Diving $\beta$ into $L$ segments of length $\Delta \tau = \beta/L$ and introducing an $\Omega$ variable for each segment, we have
\begin{equation}
    \tr(e^{-\beta H}) = \int \left[\prod_j \frac{d\Omega_j}{2 \pi} \langle \Omega_{j+1} | \Omega_{j} \rangle  \right] \exp\left(  \sum_j \Delta \tau h s_x(j) \right)=\int \left[\prod_j \frac{d\Omega_j}{2 \pi} \right] \exp\left(  \sum_j \Delta (\tau h s_x(j)+i\sin^2 \theta_j \Delta \phi_j)\right)
\end{equation}
where $s_x(i) = 3 \Omega_{x i}$. We absorb all the factors in square brackets into a measure $\CD\Omega$ to obtain the compact path integral formula
\begin{equation}
    \tr(e^{-\beta H}) = \int \CD\Omega \exp\left(  h \int_0^\beta d\tau s_x(\tau) \right).
\end{equation}

\subsection{Replicated Partition Function}

Now we give the multi-replica path integral for \name{}. For an infinitestimal imaginary time $\Delta \tau$, we have
\begin{equation}
    e^{- \Delta \tau H} = I - \Delta \tau \sum_{r_1 \mu_1 \cdots r_q \mu_q} J_{r_1 \mu_1 \cdots r_q \mu_q} \sigma_{r_1\mu_1} \cdots \sigma_{r_q \mu_q}.
\end{equation}
Using the coherent state formulas just above and the notation $s_\mu = 3 \Omega_\mu$, we have
\begin{equation}
    e^{- \Delta \tau H} = \int \prod_r \frac{d\Omega_r}{2 \pi} |\Omega_1, \cdots, \Omega_n \rangle \langle \Omega_1, \cdots, \Omega_n | \left[ 1 - \Delta \tau \sum_{r_1 \mu_1 \cdots r_q \mu_q} J_{r_1 \mu_1 \cdots r_q \mu_q} 
 s_{r_1 \mu_1} \cdots s_{r_q \mu_q} \right].
\end{equation}

We have one $\Omega$ variable for each spin and each imaginary time point, $\Omega_{ri}$. The full measure is
\begin{equation}
    \CD^N\Omega = \prod_{r} \left( \prod_i \frac{d \Omega_{ri}}{2 \pi} \langle \Omega_{ri+1} | \Omega_{ri} \rangle \right).
    \label{eq:measure_app}
\end{equation}
The path integral for the partition function of a single instance is
\begin{equation}
   Z =  \int \CD^N\Omega \exp\left( - \int d\tau \sum_{r_1 \mu_1 \cdots r_q \mu_q} J_{r_1 \mu_1 \cdots r_q \mu_q}s_{r_1 \mu_1}(\tau) \cdots s_{r_q \mu_q}(\tau) \right).
\end{equation}
We equivalently write this in terms of a multi-index $I = r_1 \mu_1 \cdots r_q \mu_q$.

The replicated path integral is obtained from the disorder average of $Z^m$, 
\begin{equation}
    \av{Z^m} = \int \CD^N\Omega_m \exp\left( \frac{\av{J_I^2}}{2 q!}\sum_{I} \left[ \sum_{a=1}^m \int_0^\beta d\tau s^a_{r_1 \mu_1} \cdots s^a_{r_q \mu_q} \right]^2 \right),
\end{equation}
where now we have included a replica index $a$ on each spin variable (and modified the measure include $m$ copies, $D\Omega_m$). The replica index runs from $1$ to $m$. The quenched free energy is obtained from the $m \to 0$ limit as discussed in the main text.

We now introduce the $G,\Sigma$ collective fields. The spin correlation $G_{ab}(\tau_1,\tau_2)$ is
\begin{equation}
    G_{ab}(\tau_1,\tau_2)=\frac{1}{N} \sum_{r\mu} s_{r \mu}^a(\tau_1) s_{r\mu}^b(\tau_2),
\end{equation}
and combined with the Lagrange multiplier $\Sigma_{ab}$, the path integral becomes
\begin{equation}
    \av{Z^m} = \int \CD^N\Omega_m \CD\Sigma \CD G \exp\left(\frac{ J^2 N}{2 q}\sum_{ab} \int G_{ab}^q - \frac{N}{2} \sum_{ab} \int \Sigma_{ab} \left[ G_{ab} - \frac{1}{N} \sum_{r\mu} s^a_{r\mu}(\tau_1) s^b_{r\mu}(\tau_2) \right] \right).
    \label{eq:path_integral_replicas}
\end{equation}
One important feature is that $G_{a\neq b}(\tau_1,\tau_2)$ should be a constant (possibly zero) since we have separate time-translation invariance in the individual replicas.

There is another representation of the path integral which decouples the $s^a s^b$ term by introducing a fluctuating magnetic field $h_{r a\mu}$ for each spin in each replica. The fields are independent between spins, and for a single spin across all replicas, the covariance of $h_{a\mu}(\tau_1)$ and $h_{b\nu}(\tau_2)$ is $\delta_{\mu \nu} \Sigma_{ab}(\tau_1,\tau_2)$. This includes the possibility that $h_{a\mu}(\tau)$ is correlated between replicas. The path integral is then expressed as
\begin{equation}
    \av{Z^m} = \int \CD^N\Omega_m \CD\Sigma \CD G \CD^N h \exp\left(\frac{ J^2 N}{2 q}\sum_{ab} \int G_{ab}^q - \frac{N}{2} \sum_{ab} \int \Sigma_{ab}  G_{ab}  + \sum_{ra} \int d\tau h_{r a \mu}(\tau) s^a_{r\mu}(\tau) \right),
\end{equation}
where $\CD^N h$ is a product of Gaussian measures for $h_{ra\mu}$. If we now do the $\Omega$ and $h$ integrals, we can write the replicated partition function as
\begin{equation}
    \av{Z^m} = \int \CD\Sigma \CD G \exp\left(\frac{ J^2 N}{2 q}\sum_{ab} \int G_{ab}^q - \frac{N}{2} \sum_{ab} \int \Sigma_{ab}  G_{ab}  + N \log \langle \mathcal{P} e^{\int d\t h_a \cdot s_a} \rangle_{\Sigma_{ab}} \right),\label{eq:ssyk_rep_action}
\end{equation}

The single replica path integral is obtained by setting $m=1$. It is 
\begin{equation}
     \av{Z} = \int \CD^N \Omega \CD\Sigma \CD G \exp\left(\frac{ J^2 N}{2 q} \int G^q - \frac{N}{2} \int \Sigma \left[ G - \frac{1}{n} \sum_{r\mu} s_{r\mu}(\tau_1) s_{r\mu}(\tau_2) \right] \right)
\end{equation}
or, equivalently,
\begin{equation}
    \av{Z} = \int \CD^N\Omega \CD\Sigma \CD G \CD^N h \exp\left(\frac{ J^2 N}{2 q} \int G^q - \frac{N}{2}  \int \Sigma  G  + \sum_{r} \int d\tau h_{r \mu}(\tau) s_{r\mu}(\tau) \right).
\end{equation}




\section{Numerical methods}
\label{app:num_methods}

\subsection{Solving the Equations of Motion }

For the data in Sec.~\ref{sec:Numerics}, we used MATLAB to setup an iterative solver scheme on a discretized imaginary time circle. The solver has a few components.

First, given a value of $\Sigma$, we need to obtain an estimate of $G$. As stated in the main text, we use sampling to do this. However, we want to do importance sampling from the full distribution (single spin partition function times the Gaussian distribution) of the magnetic fields rather than just randomly sampling from the Gaussian distribution. To do this, we implemented a Markov chain Monte Carlo based on the Metropolis-Hastings algorithm. We generated candidate steps in the magnetic field by sampling from the Gaussian distribution and multiplying by a small factor, typically of order $.1$ or $.2$. We then randomly accept or reject the step based on the Metropolis-Hastings rule. After some number of steps, typically $6$, we use the resulting fields to generate one sample contributing to $G$. We do not know the auto-correlation time of our Markov chain, so these parameters are largely ad hoc, chosen after much experimentation to give reliable solutions.

Second, given a value of $G$, we need to obtain the corresponding $\Sigma$. This is trivial using the $G$ EOM. We do enforce that $G$ is reflection symmetric in the above sampling procedure, and we also enforce that $\Sigma$ is symmetric and translation invariant.

Third, we need an update rule. We choose a mix parameter, $x$ and a fraction $b$,  and propose updates via
\begin{equation}
    G_{\text{prop}} = (1-x) G_{\text{old}} + x G_{\text{new}}
\end{equation}
where $G_{\text{new}}$ is obtained from the sampling subroutine above. There is an identical update rule for $\Sigma$. If this update does not increase the error, then we accept it and continue. If it does increase the error, then we reduce $x \to bx$ and continue. We always carry out at least one update since the error is initialized to a large value. Typically, $b=.5$ and $x$ is initialized to $.5$.

The data in the main text allows for $6$ iterations of this procedure starting from an initial guess. When not using enough time points or enough samples, one often finds that no updates after the first are accepted. However, with sufficient time points and samples, we have found that nearly every update can be accepted.

The number of iterations and time points are quite modest compared to what is standard for solutions of the SYK equations of motion. However, the sampling procedure is currently very expensive. The most expensive runs, e.g. Figure~\ref{fig:q3_eom_syk}, with $N_\tau=300$ time points and $N_s=10^4$ samples took around $5$ hours to run on a single core. The code was run on the Zaratan cluster at University of Maryland, College Park.

We are not sure if there is a vastly more efficient method for solving the equations of motion. However, at least by parallelizing the sampling and making some other modest improvements, one could probably study significantly larger $N_\tau$ and $N_s$.

\subsection{Finite Size Numerics}

For the data in Sec.~\ref{sec:ed}, we used MATLAB and defined instances of the sparse \name{} model as sparse matrices. We then made calls to MATLAB's built-in sparse matrix eigenvalue routines to access the lowest few eigenvalues and eigenvectors of the sparse Hamiltonian. The code was run on the Zaratan cluster at University of Maryland, College Park.

\bibliographystyle{ieeetr}
\bibliography{main.bib}

\end{document}